\documentclass[tightenlines,showpacs,prd,floatfix,preprintnumbers,amsmath,amssymb,nofootinbib,superscriptaddress]{revtex4}
\DeclareMathAlphabet{\mathpzc}{OT1}{pzc}{m}{it}
\usepackage{stmaryrd}
\usepackage{graphicx}% Include figure filesg
\usepackage{amsfonts}
\usepackage[normalem]{ulem}
\usepackage{color}
\usepackage{bm}
\usepackage{comment}
\usepackage{ulem}
\def\pheq{\phantom{=}}
\def\up#1{\raise1mm\hbox{$\!\!^{#1}$}} 
\def\upp#1{\raise2mm\hbox{$\!\!\!\!^{#1}$}}

\newcommand{\beq}{\begin{equation}}
\newcommand{\be}{\begin{equation}}
\newcommand{\eeq}{\end{equation}}
\newcommand{\ee}{\end{equation}}
\newcommand{\bea}{\begin{eqnarray}}
\newcommand{\eea}{\end{eqnarray}}
\newcommand{\bal}{\begin{align}}
\newcommand{\eal}{\end{align}}

\newcommand{\beaa}{\begin{eqnarray*}} 
\newcommand{\eeaa}{\end{eqnarray*}} 
\newcommand{\pa}{\partial}
\newcommand{\na}{\nabla}
\newcommand{\dis}{\displaystyle} 
\newcommand{\Lie}{\mbox{\pounds}}
\newcommand{\bsube}{\begin{subequations}}
\newcommand{\esube}{\end{subequations}}

\newcommand{\preto}{\psi_0^{\rm ret}}

\newcommand{\somucho}{\frac{im\left[ a-\Omega\left( r_0^2+a^2\right)\right]}{\Delta_0}}

\begin{document}

\title{EMRI corrections to the angular velocity and redshift factor of a mass in circular orbit about a Kerr black hole}

\author{Abhay G. Shah} 
\email{abhay.shah@weizmann.ac.il}
\affiliation{Dept of Particle Physics $\&$ Astrophysics, 
Weizmann Institute of Science, Rehovot 76100, Israel}
\affiliation{Center for Gravitation and Cosmology, Department of Physics, 
University of Wisconsin--Milwaukee, P.O. Box 413, Milwaukee, Wisconsin 53201, 
USA}

\author{John L. Friedman} 
\email{friedman@uwm.edu}
\affiliation{Center for Gravitation and Cosmology, Department of Physics, 
University of Wisconsin--Milwaukee, P.O. Box 413, Milwaukee, Wisconsin 53201, 
USA}

\author{Tobias S. Keidl} 
\email{tobias.keidl@uwc.edu}
\affiliation{Department of Physics, 
University of Wisconsin--Washington County,USA}

\date{2012}

\begin{abstract}
This is the first of two papers on computing the self-force in a radiation gauge for a particle of mass $\frak{m}$ moving in circular, equatorial orbit about a Kerr black hole.  In the EMRI (extreme-mass-ratio inspiral) framework, 
with mode-sum renormalization, we compute 
the renormalized value of the quantity $H:=\frac12 h_{\alpha\beta}u^\alpha u^\beta$, gauge-invariant under gauge transformations generated by a helically symmetric gauge vector; 
here $h_{\alpha\beta}$ is the metric perturbation,  $u^\alpha$ the particle's 4-velocity. We find the related order $\frak{m}$ correction to the particle's angular velocity at fixed renormalized redshift (and to its redshift at fixed angular velocity), each of which can be written in terms of $H$.   The radiative part of the metric perturbation is constructed from a Hertz potential that is extracted from the Weyl scalar by an algebraic inversion\cite{sf2}. We then write the spin-weighted spheroidal harmonics as a sum over spin-weighted spherical harmonics $_sY_{\ell m}$ and use mode-sum renormalization to find the renormalization coefficients by matching a series in $L=\ell+1/2$ to the large-$L$ behavior of the expression for $H$. The non-radiative parts of the perturbed metric associated with changes in mass and angular momentum are calculated in the Kerr gauge.
\end{abstract}

\maketitle

\section{Introduction}

We report here a first computation of the order $\frak{m}$ renormalized corrections $\Delta \Omega$ to the angular velocity of a mass $\mathfrak m$ in circular orbit about a Kerr black hole and of the related change $\Delta U$ 
in the scalar $U=u^\alpha \nabla_\alpha t$.  This is the constant of proportionality between the particle's velocity 
and the helical Killing vector $k^\alpha$ tangent to its trajectory: $u^\alpha = U k^\alpha$.   
The corresponding computation for a Schwarzschild background was first done by Detweiler \cite{detweiler08}, 
who pointed out the invariance of these quantities under gauge transformations generated by helically symmetric gauge vectors.  The invariance allows straightforward comparison with computations in other gauges, and comparisons were done for 
circular orbits in Schwarzschild for a Lorenz gauge \cite{sbd08} and for our modified radiation gauge \cite{sf3}.  
Previous conservative self-force computations in Kerr spacetime have been restricted to a scalar particle 
in eccentric, equatorial orbit \cite{WB-1,WB-2}.

The radiation gauges are associated with the Weyl scalars $\psi_0$ and $\psi_4$ that satisfy the 
Teukolsky equation \cite{1973ApJ...185..635T}: They are gauges in which one can construct a perturbed 
metric in the vacuum spacetime outside the particle from $\psi_0$ or $\psi_4$, and for circular orbits 
this can be done by an algebraic inversion.  The resulting metric, however, lacks the non-radiative contributions 
from the changes $\delta M$ and $\delta J$ in mass and angular momentum, and from the change in the 
center of mass.  We write the metric perturbations corresponding to $\delta M$ and $\delta J$ in the 
Kerr gauge, the result of changing $a$ and $M$ in the Kerr metric written in Boyer-Lindquist coordinates.  It is in this 
modified radiation gauge that we compute $\Delta\Omega$ and $\Delta U$. 

We believe that the numerical results given here are accurate to better
than one part in $10^{8}$, consistent with our earlier Schwarzschild
computation.  Numerical values we reported in an
earlier eprint version of this paper had significant inaccuracy, due
to an oversight in our computation of the singular field,
found after two comparisons.  We performed a detailed comparison of
our results for $H^{\rm ren}$ with corresponding post-Newtonian values found by
A. Le Tiec; and we compared a few values of $H^{\rm ren}$ with those of
S. Dolan, who has completed another EMRI calculation
of $H^{\rm ren}$.  Dolan's calculation uses a Lorenz gauge and
an effective-source renormalization method (see, e.g. \cite{bgs07,vd08,vwd11,dbw11}).
Where our initial error led to a numerical error of about 5\%
(with about twice that error for the the difference
between $H$ for Schwarzschild and for Kerr),  Dolan's present
accuracy is about 1\%. 
\footnote{
Dolan's results are to be reported in a paper in preparation 
coauthored by L. Barack and B. Wardell.}

The plan of the paper is as follows. In Sec.~\ref{sec2a}, after 
a brief review of the Teukolsky equation and associated formalism, we obtain the analytic form of the Weyl scalar $\psi_0^{\rm ret}$, 
expressed as a sum over spin-weighted spheroidal harmonics. 
In \ref{sec:hrad} we review the construction of the metric perturbation 
in a radiation gauge, using a Hertz potential constructed 
algebraically from $\psi_0$.  We describe the mode sum renormalization,
pointing out a difference, not present in our Schwarzschild 
computation, between the singular field at subleading order in a Lorenz 
gauge and in our radiation gauge. 
In Sec. \ref{sec2c}, we first review two different series expansions 
for spin-weighted spheroidal harmonics, one as a sum of Jacobi polynomials, the other as a sum over spin-weighted spherical harmonics. 
We check the accuracy of the angular harmonics by computing them using these two different expansions.
We then obtain explicit expressions for the values of the harmonics  
$_1Y_{\ell m}$ and $_2 Y_{\ell m}$ at $\theta = \pi/2$ (the plane 
of the particle's orbit).  In Sec. \ref{sec2d} we review the 
numerical construction of solutions to the radial Teukolsky equation, 
by direct numerical integration and by numerical integration of 
the Sasaki-Nakamura equation.    
Again we check the accuracy of our computation of the radial harmonics and the angular harmonics of $\psi_0^{\rm ret}$ obtained in these 
two independent ways.   
In Sec. \ref{sec4} we compute the tetrad components of the perturbed metric from the Hertz potential. In Sec. \ref{sec5} we find expressions for 
the change $\Delta U$ in the renormalized redshift factor at fixed $\Omega$ and the corresponding change in the angular 
velocity at fixed $U$; each is simply related to the quantity 
$\dis H^{\rm ren} =\frac12 h^{\rm ren}_{\alpha\beta} u^\alpha u^\beta$.
In Sec. \ref{sec6} we present the computation of the retarded component of the lower multipoles in Boyer-Lindquist coordinates. In Sec. \ref{sec7}, we present the numerical results for $H^{\rm ren}$, $\Delta U$ and $\Delta \Omega$.  Finally, in Sec. VII, we 
briefly discuss our results and future work.

\section{Review and computation of $\preto$}
\label{sec2}

\subsection{Formalism}
\label{sec2a}

We work in Boyer-Lindquist coordinates where the Kerr metric is given by 
\begin{align}
ds^2 &= \left( 1-\frac{2Mr}{\Sigma}\right)dt^2 + \frac{4Mar\sin^2\theta}{\Sigma}dtd\phi - \frac{\Sigma}{\Delta}dr^2 - \Sigma d\theta^2 - \left( r^2+a^2+\frac{2Mra^2\sin^2\theta}{\Sigma}\right)\sin^2\theta d\phi^2,
\end{align}
with $\Delta = r^2-2Mr+a^2$ and $\Sigma = r^2+a^2\cos^2\theta$.
The Kinnersley tetrad vectors have components 
\begin{align}
l^\alpha &= \frac{1}{\Delta}\left( r^2+a^2, \Delta, 0, a\right), \nonumber \\
n^\alpha &= \frac{1}{2\Sigma}\left( r^2+a^2, -\Delta, 0, a\right), \nonumber \\
m^\alpha &= \frac{-\bar{\varrho}}{\sqrt{2}}\left(ia\sin\theta, 0, 1, \frac{i}{\sin\theta} \right),
\label{eq:tetrad}\end{align}
and we denote by ${\bm{D}}, {\bm{\Delta}}$ and ${\bm{\delta}}$ the derivative operators along the tetrad vectors $l^\alpha, n^\alpha$ and $m^\alpha$, respectively. The non-vanishing spin coefficients associated with this tetrad are
\begin{align}
\varrho &= \frac{-1}{r-ia\cos \theta}, \quad \beta = \frac{-\bar{\varrho} \cot\theta}{2\sqrt{2}}, \quad \varpi = \frac{ia\varrho^2\sin\theta}{\sqrt{2}}, \quad \tau = \frac{-ia\sin\theta}{\sqrt{2}\Sigma}, \quad \mu = \frac{\Delta \varrho}{2\Sigma}, \quad \gamma = \mu + \frac{(r-M)}{2\Sigma}, \quad \alpha = \varpi - \bar{\beta}.
\label{eq:spincoeff}\end{align}

Consider a particle of mass $\frak{m}$ orbiting a Kerr black hole in circular, equatorial orbit with radial coordinate 
$r=r_0$. As noted in the introduction, the particle's velocity is tangent to a helical Killing vector, 
\beq
u^\alpha = u^t k^\alpha, 
\label{eq:ut}\eeq
where 
\be
k^\alpha = t^\alpha + \Omega \phi^\alpha,
\label{eq:k}\ee 
with $\phi^\alpha$ and $t^\alpha$ the rotational and asymptotically timelike Killing vectors of 
the Kerr geometry and $\Omega$ the particle's angular velocity measured by an observer at infinity.  
For a circular geodesic, the values of $u^t$ and $\Omega$ are 
\bea\label{utOmega}
u^t &=& \frac{r_0^{3/2}\pm M^{1/2}a}{\sqrt{r_0^3-3Mr_0^2\pm2aM^{1/2}r_0^{3/2}}}, \\ \nonumber
\Omega &=& \frac{\pm M^{1/2}}{r_0^{3/2} \pm a M^{1/2}},
\eea
where the upper (lower) sign correspond to direct (retrograde) orbits and the corresponding stress-energy tensor is given by
\begin{align}
T^{\alpha\beta} = \frac{\frak m}{u^tr_0^2}u^\alpha u^\beta \delta(r-r_0)\delta(\theta - \pi/2)\delta(\phi-\Omega t).
\label{stress}\end{align}
The perturbed spin-2 Weyl scalar, $\psi_0= -C_{\alpha\beta\gamma\delta}l^\alpha m^\beta l^\gamma m^\delta$,  satisfies the Teukolsky equation
\begin{align} \label{Teuk_in_Kerr}
&\left[ \frac{(r^2+a^2)^2}{\Delta}-a^2\sin^2\theta\right]\frac{\partial^2\psi_0}{\partial t^2} + \frac{4Mar}{\Delta}\frac{\partial^2\psi_0}{\partial t\partial\phi} + \left[ \frac{a^2}{\Delta}-\frac{1}{\sin^2\theta}\right]\frac{\partial^2\psi_0}{\partial\phi^2} - \frac{1}{\Delta^2}\frac{\partial}{\partial r}\left( \Delta^3 \frac{\partial \psi_0}{\partial r}\right) - \frac{1}{\sin\theta}\frac{\partial}{\partial\theta}\left( \sin\theta \frac{\partial\psi_0}{\partial\theta}\right) \nonumber \\
&- 4\left[\frac{a(r-M)}{\Delta} + \frac{i\cos\theta}{\sin^2\theta}\right]\frac{\partial\psi_0}{\partial\phi} + 2(2\cot^2\theta-1)\psi_0 - 4\left[\frac{M(r^2-a^2)}{\Delta}-r-ia\cos\theta\right]\frac{\partial\psi_0}{\partial t}= 8\pi\Sigma T,
\end{align}
whose source $T$ is given by 
\begin{align}
T &:= - \left( {\bm \delta} + \bar{\varpi} - \bar{\alpha} - 3\beta - 4\tau \right)\left( {\bm \delta} + \bar{\varpi} - 2\bar{\alpha} - 2\beta \right)T_{\bm{11}} \nonumber \\
&\quad\, + \left[\left( {\bm \delta} + \bar{\varpi} - \bar{\alpha} - 3\beta-4\tau\right)\left({\bm D}-2\bar{\varrho}\right) + \left({\bm D} - 4\varrho - \bar{\varrho}\right)\left({\bm \delta} + 2\bar{\varpi} - 2\beta\right)\right] T_{\bm{13}} \nonumber \\
&\quad\, - \left( {\bm D} - 4\varrho - \bar{\varrho}\right)\left( {\bm D} - \bar{\varrho} \right) T_{\bm{33}} \nonumber \\ 
&=  T^{(0)} + T^{(1)} + T^{(2)}, 
\label{eq:tsource}\end{align}
where the $T^{(i)}$ label the three terms in the expression for $T$. 
 
Because the Teukolsky equation is separable in the frequency domain, we can write $\psi_0$ as a 
sum of angular harmonics, 
\beq \label{psi0decomp}
\psi_0 = \sum_{\ell, m}R_{\ell m}(r) {\,}_2 S_{\ell m}(\theta)e^{im(\phi-\Omega t)},
\eeq
where we used the fact that, for circular orbits, harmonics $e^{im\phi}$ 
have frequency $\omega = m\Omega$. 
For $r \neq r_0$, the source vanishes, and $R_{\ell m}$ satisfies the radial equation
\beq
\Delta R^{\prime \prime}_{\ell m} + 6(r-M)R^\prime_{\ell m} + \left( \frac{K^2-4iK(r-M)}{\Delta}+8im\Omega r - \gamma^2+2m\gamma+6-{\,}_2E_{\ell m}\right)R_{\ell m}=0.
\label{eq:teukr}\eeq
The function ${\,}_2S_{\ell m}$ satisfies the angular equation
\beq \label{2Slm}
\frac{1}{\sin\theta}\frac{d}{d\theta}\left(\sin\theta\frac{d}{d\theta}\ {}_2S_{\ell m}\right)
         + \left( \gamma^2\cos^2\theta-4\gamma\cos\theta 
        - \frac{4m\cos\theta+4+m^2}{\sin^2\theta}+{\,}_2E_{\ell m} \right){}_2S_{\ell m} = 0,
\eeq
where $K=m\Omega(r^2+a^2)-am$ and $\gamma = am\Omega$.

The tetrad components of the stress-energy tensor that enter the expression for $T$ in Eq. (\ref{Teuk_in_Kerr})  are
\begin{align}
T_{\bm {11}} &= \frac{{\frak m}(1-a\Omega)^2u^t}{r_0^2}\delta(r-r_0)\delta(\theta - \pi/2)\delta(\phi-\Omega t), \nonumber \\
T_{\bm{13}} &= \frac{i{\frak m}(1-a\Omega)\left(a-\Omega(r_0^2+a^2)\right)u^t}{\sqrt{2}r_0^3}\delta(r-r_0)\delta(\theta - \pi/2)\delta(\phi-\Omega t), \nonumber \\
T_{\bm{33}} &= \frac{-{\frak m} \left(a-\Omega(r_0^2+a^2)\right)^2u^t}{2r_0^4}\delta(r-r_0)\delta(\theta - \pi/2)\delta(\phi-\Omega t).
\end{align}
Using Eqs.~(\ref{eq:tetrad}) to write the derivative operators and Eqs.~(\ref{eq:spincoeff}) for 
the spin coefficients, we obtain explicit forms for the source terms $T^{(i)}$ of Eq.~(\ref{eq:tsource}). 
Defining the quantities 
\be 
 q := \frac{m\left[ a-\Omega\left( r^2+a^2\right)\right]}{\Delta},
\qquad q_0 := q|_{r=r_0} 
= \frac{m\left[ a-\Omega\left( r_0^2+a^2\right)\right]}{\Delta_0},
\label{eq:q}\ee
and writing 
\be
   \delta(\phi-\Omega t) = \sum_{m=-\infty}^\infty \frac1{2\pi}e^{im(\phi-\Omega t)}, 
\ee
we find 
\begin{align}
T^{(0)} &=  -\frac1{4\pi}{\mathfrak m}\sum_{m=-\infty}^{\infty}\frac{\bar{\varrho}^2}{r_0^2}
  (1-a\Omega)^2u^t e^{im(\phi-\Omega t)}\Biggl\{ \delta^{\prime\prime}(\theta - \pi/2)  \nonumber \\
&  \qquad\qquad + \left[ m a\Omega(1+\sin\theta) - m-\frac{m}{\sin\theta} - \cot\theta +\frac{ia}{r_0} 
 - 4ia\varrho \sin\theta  - ia\bar{\varrho}\sin\theta\right]\delta^{\prime}(\theta - \pi/2) \nonumber \\
& \qquad\qquad+ \left[m(1-a\Omega)-i\frac{a}{r_0}\right]\left[m(1-a\Omega)-5i\frac{a}{r_0}\right]\delta(\theta - \pi/2)\Biggr\}\delta(r-r_0) ,
\end{align}
\begin{align}
T^{(1)} &= -\frac i{4\pi}{\mathfrak m}\sum_{m=-\infty}^{\infty}\frac{\bar{\varrho}}{r_0^3} (1-a\Omega)\left[a-\Omega(r_0^2+a^2)\right]u^te^{im(\phi-\Omega t)}\left\{  2\delta^\prime(r-r_0)\delta^\prime(\theta-\pi/2) \phantom{\frac12}\right.\nonumber \\
& \qquad\qquad + 2\left(\frac{1}{r_0} - 2\varrho +iq_0\right)\delta(r-r_0)\delta^\prime(\theta-\pi/2) 
%\nonumber \\
+ \left( 2am\Omega - 2m + \frac{4ia}{r} - \frac{2ia}{r_0}\right)\delta^\prime(r-r_0)\delta(\theta-\pi/2) \nonumber \\
&\qquad\qquad + \left[ \left(am\Omega-m+\frac{4ia}{r_0}\right)\left(\frac{2}{r_0}+iq_0\right)   
%\nonumber \\ &
  \left. + \left(am\Omega-m-\frac{2ia}{r_0}\right) \left(\frac{4}{r_0}+iq_0\right)\right] \delta(r-r_0)\delta(\theta-\pi/2)\right\}, 
\end{align}
\begin{align}
T^{(2)} &= -\frac1{4\pi}{\frak m}\sum_{m=-\infty}^{\infty}\frac1{r_0^4} \left[ a-\Omega\left(r_0^2+a^2\right)\right]^2 u^t e^{im(\phi-\Omega t)}\left\llbracket -\delta^{\prime\prime}(r-r_0) 
\phantom{\frac 1{\frac12}}\right.\nonumber \\
& \hspace{2.3cm}
 - \left\{ i(q+q_0) + \frac{1}{r_0} + \frac{5}{r}\right\}\delta^\prime(r-r_0) - \left. \left( \frac{5}{r_0}+iq_0\right) \left(\frac{1}{r_0}+iq_0\right) \delta(r-r_0)\right\rrbracket \delta(\theta-\pi/2).
\end{align}

To compute $\psi_0^{\rm ret}$, it is useful to define a Green's function $G_{\ell m}(r,r^\prime)$ as a solution to
\begin{align} 
\Delta G^{\prime \prime}_{\ell m} + 6(r-M)G^\prime_{\ell m} + \left[ \frac{K^2-4iK(r-M)}{\Delta}+8im\Omega r - \gamma^2+2m\gamma+6-{\,}_2E_{\ell m}\right]G_{\ell m}= \delta(r-r^\prime),
\end{align}
namely
\begin{align}
G_{\ell m}(r,r^\prime) &= A_{\ell m} \Delta^{\prime^2}R_H(r_<)R_\infty(r_>), \\
{\rm where} \quad  A_{\ell m} &:= \frac{1}{\Delta^3(R_H R_\infty^\prime - R_\infty R_H^\prime)}.
\end{align}
Here $R_H$ and $R_\infty$ are two linearly independent solutions to the radial equation that are, respectively,
ingoing at the future horizon and outgoing at future null infinity, and we have suppressed 
the $\ell m$ indices on $R_H$ and $R_\infty$. 
With $G_{\ell m}$ so defined, the full Green's function that satisfies the Teukolsky equation with source 
$\delta(r-r')\delta(\cos\theta-\cos\theta')\delta(\phi-\phi')$  is given by
\begin{align}
G(x,x^\prime) = \sum_{\ell m}A_{\ell m}\Delta^{\prime^2}R_H(r_<)R_\infty(r_>){\,}_2S_{\ell m}(\theta){\,}_2S_{\ell m}(\theta^\prime)e^{im(\phi-\phi^\prime)},
\end{align}
where we denote by $x$ the spatial point with coordinates $r,\theta,\phi$ and where $\Delta^\prime = r^{\prime^2}-2Mr^\prime+a^2$.    
Hence, the Weyl scalar that satisfies the Teukolsky equation has the form
\begin{align}
\psi_0 &= -8\pi\int\Sigma^\prime T(x^\prime,x_0)G(x,x^\prime)dr^\prime d(\cos\theta^\prime)d\phi^\prime \nonumber \\
&= -8\pi\int\Sigma^\prime (T^{(0)} + T^{(1)} + T^{(2)}) G(x,x^\prime)dr^\prime d(\cos\theta^\prime)d\phi^\prime \nonumber \\
&=: \psi_0^{(0)} + \psi_0^{(1)} + \psi_0^{(2)}
\end{align}
with $\Sigma^\prime = r^{\prime^2}+a^2\cos^2\theta^\prime$.  The $\psi^{(i)}$ have, for $r\neq r_0$, the explicit forms,
\begin{align}
\psi_0^{(0) }&=\frac{4\pi{\frak m}(1-a\Omega)^2\Delta_0^
 2u^t}{r_0^2}\sum_{\ell m}A_{\ell m}R_H(r_<)R_\infty(r_>){\,}_2S_{\ell m}(\theta)e^{im(\phi-\Omega t)}\nonumber \\
&\quad \Biggl[ S_0^{\prime\prime} + \frac{2iaS_0^\prime}{r_0} - 2S_0 + 2m(a\Omega-1)S_0^\prime + \frac{2iam(a\Omega-1)S_0}{r_0}  + m^2(a\Omega-1)^2S_0\Biggr],\\
%\end{align}
%\begin{align}
\psi_0^{(1)} &=\frac{8\pi i{\frak m}(1-a\Omega)\left[a-\Omega(r_0^2+a^2)\right]u^t}{2r_0^3}\sum_{\ell m}A_{\ell m}{\,}_2S_{\ell m}(\theta)e^{im(\phi-\Omega t)} \nonumber \\ 
& \qquad\left\llbracket 2\Delta_0^2\left[ r_0S_0^\prime-iaS_0 + r_0S_0\left(am\Omega-m+\frac{ia}{r_0}\right)\right] 
\left[ R_H^\prime(r_0)R_\infty(r)\Theta(r-r_0) + R_\infty^\prime(r_0)R_H(r)\Theta(r_0-r)\right]\right. \nonumber \\ 
%\displaybreak 
&\qquad + \Biggl\{2\Delta_0\left[S_0^\prime(5r_0^2-6Mr_0+a^2)-4iaS_0(r_0-M)\right] -2\Delta_0^2(r_0S_0^\prime-iaS_0) \left[\somucho+\frac{3}{r_0}\right] \nonumber \\
&\hskip14mm- S_0r_0\Delta_0^2\left[ \left(am\Omega-m+\frac{4ia}{r_0}\right)\left(\somucho+\frac{2}{r_0}\right) \right. \nonumber \\
& \left. \hskip30mm+\left(\somucho+\frac{4}{r_0}\right)\left(am\Omega-m - \frac{2ia}{r_0}\right)\right] \nonumber \\
& \left.\hskip14mm + 2\Delta_0S_0\left(am\Omega-m+\frac{ia}{r_0} \right)(5r_0^2-6Mr_0+a^2) - \frac{8iaS_0\Delta_0^2}{r_0}\Biggr\} R_H(r_<)R_\infty(r_>) \right\rrbracket,\\
%\end{align}
%
%\begin{align}
\psi_0^{(2)} &=  \frac{4\pi{\frak m} u^t \left(a-\Omega(r_0^2+a^2)\right)^2}{r_0^4}\sum_{\ell m}A_{\ell m}{\,}_2S_{\ell m}(\theta){\,}_2S_{\ell m}(\pi/2)e^{im(\phi-\Omega t)} \times 
\nonumber \\
& \qquad \left\llbracket r_0^2 \Delta_0^2\left\{R_H^{\prime\prime}(r_0)R_\infty(r) \Theta(r-r_0) + R_\infty^{\prime\prime}(r_0)R_H(r)\Theta(r_0-r)\right\} \phantom{\frac1{\frac12}}\right.\nonumber \\
&\qquad\quad + \left\{ 4r_0\Delta_0(3r_0^2-4Mr_0+a^2)-2r_0^2\Delta_0^2\left[\somucho + \frac{3}{r_0}\right]\right\}\times \nonumber \\
&\hspace{3cm} \left\{ R_H^\prime(r_0)R_\infty(r)\Theta(r-r_0) + R_\infty^\prime(r_0)R_H(r)\Theta(r_0-r)\right\} \nonumber \\
&\qquad\quad + \left\{ r_0^2\Delta_0^2\left[\somucho+\frac{5}{r_0}\right]\left[\somucho + \frac{1}{r_0}\right] \right. \nonumber \\
& \qquad\left. \qquad - r_0^2\Delta_0^2\left[\frac{-5}{r_0^2} - \frac{2im\Omega r_0}{\Delta_0}-\frac{2im(r_0-M)(a-\Omega(r_0^2+a^2))}{\Delta_0^2}\right] \right. \nonumber \\
&\qquad\left. \qquad - 4r_0\Delta_0(3r_0^2-4Mr_0+a^2)\left[\somucho + \frac{3}{r_0}\right] \right. \nonumber \\
&\qquad\left. \qquad + 2(3r_0^2-4Mr_0+a^2)^2 + 4r_0\Delta_0(3r_0-2M)\Biggr\} R_H(r_<)R_\infty(r_>)
\right\rrbracket,
\end{align}
where $\Delta_0 = r_0^2-2Mr_0+a^2$, $S_0 = {\,}_2S_{\ell m}(\pi/2)$; $S_0^\prime$ and $S_0^{\prime\prime}$ are derivatives of ${\,}_2S_{\ell m}$ with respect to $\cos\theta$ evaluated at $\theta = \pi/2$.

\subsection{Metric perturbation and $H^{\rm ren}$ in a radiation gauge}
\label{sec:hrad}

Before presenting the detailed calculation of $H$, $\Delta U$ and $\Delta\Omega$, it will 
be helpful briefly to recall parts of the radiation-gauge method that we will use.  In the CCK procedure 
\cite{chr75,ck74,kc79,stewart79,wald78} a perturbed vacuum metric is constructed from a 
spin $2$ Hertz potential $\Psi$ that satisfies 
\bea
\psi_0 = \frac{1}{8}\left[ {\cal L}^4 \bar{\Psi} + 12M\partial_t \Psi\right],
\eea
where ${\cal L}^4 = {\cal L}_1 {\cal L}_0 {\cal L}_{-1} {\cal L}_{-2}$ with ${\cal L}_s = \eth_s - ia\sin\theta\partial_t$ and $\eth_s = -\left[ \partial_\theta + i\csc\theta\partial_\phi - s\cot\theta \right]$.
Using Eq (\ref{psi0decomp}) and Eq (40) of \cite{sf2}, we obtain
\bea \label{Psieqn}
\Psi &=& 8\sum_{\ell m}\frac{(-1)^mD\bar{R}_{\ell m}+12imM\Omega R_{\ell m}}{D^2+144M^2m^2\Omega^2}e^{im(\phi-\Omega t)}{\,}_2S_{\ell m}(\theta),
\eea
where $D^2=\lambda_{CH}^2(\lambda_{CH}+2)^2+8a\omega(m-a\omega)\lambda_{CH}(5\lambda_{CH}+6) + 48a^2\omega^2[2\lambda_{CH}+3(m-a\omega)^2]$ 
and $\lambda_{CH}$, the angular eigenvalue used by Chandrasekhar \cite{Chandra},  is related to the separation constant ${\,}_2E_{\ell m}$ 
of Eq.~(\ref{eq:teukr}) by $\lambda_{CH}={\,}_2E_{\ell m} +\gamma^2-2m\gamma - 2$. 
The perturbed metric is then given (up to parts for which $\psi_0$ vanishes) by  
\bea
h_{\alpha\beta}&=&\varrho^{-4}\{\, n_\alpha n_\beta (\bar{\bm\delta}-3\alpha-\bar{\beta} +5\pi)(\bar{\bm\delta}-4\alpha+\pi)+\bar{m}_\alpha\bar{m}_\beta(\bm\Delta+5\mu-3\gamma + \bar \gamma)(\bm\Delta+\mu-4\gamma) \nonumber\\
&\pheq& - n_{(\alpha}\bar{m}_{\beta)} \left[(\bar{\bm \delta} - 3\alpha +\bar \beta +5\pi + \bar \tau) (\bm \Delta + \mu - 4\gamma)+
(\bm \Delta + 5\mu -\bar \mu -3\gamma -\bar \gamma)(\bar{\bm \delta} -4\alpha + \pi) \right]\, \} \Psi+\rm{c.c.}.
\label{eq:hab}
\eea

% Our computation of $H^{\rm ren}_\ell$ uses an additional feature of the renormalized 
% metric that underlies a more general description of renormalization in a radiation gauge.
To describe the way one renormalizes the metric in a radiation gauge, it is helpful to 
begin with the Detweiler-Whiting \cite{Detweiler:2002mi} decomposition of 
the metric in a Lorenz gauge.  This has the form 
$h^{\rm ret, Lor}_{\alpha\beta} = h^{\rm S, Lor}_{\alpha\beta} +  h^{\rm ren, Lor}_{\alpha\beta}$, with  $h^{\rm ren, Lor}_{\alpha\beta}$ a smooth solution to the vacuum Einstein 
equation in a neighborhood of the particle.  Because the perturbed Weyl scalar $\psi_0$ is 
gauge invariant, when computed from this decomposed form of $h^{\rm ret, Lor}_{\alpha\beta}$, 
it has the gauge-invariant decomposition $\psi_0^{\rm ret} = \psi_0^S + \psi_0^{\rm ren}$, 
with $\psi_0^{\rm ren}$ a smooth sourcefree solution to 
the $s=2$ Teukolsky equation.   There is then a smooth
Hertz potential $\Psi^{\rm ren}$ satisfying the sourcefree $s=2$ Teukolsky equation 
with $h^{\rm ren, ORG}_{\alpha\beta}$ given in terms of $\Psi^{\rm ren}$ by Eq.~(\ref{eq:hab}).  
Here $h^{\rm ren, ORG}_{\alpha\beta}$ is the reconstructed perturbed metric up to terms involving 
changes in the mass, angular momentum, and center of mass.  (These are metric perturbations 
for which $\psi_0$ vanishes, and they correspond to the $\ell = 0$ and $\ell=1$ 
parts of the perturbation for a Schwarzschild background.)  Adding 
these perturbations in an arbitrary gauge that is smoothly related to the 
Lorenz gauge yields the full perturbed metric in a modified 
radiation gauge.  

We will use this general description at the end of Sec.~\ref{sec4}, but
in renormalizing $H^{\rm ret} = \frac12 h^{\rm ret}_{\alpha\beta}u^\alpha u^\beta$, 
we exploit its invariance under gauge transformations generated by helically symmetric gauge vectors. 
As in our Schwarzschild paper \cite{sf3}, this allows 
us to use the generic Lorenz-gauge singular behavior of the metric perturbation for the leading term in the singular field.
A difference between the singular field in our radiation-gauge and 
in the Lorenz-gauge arises at subleading order in the angular harmonic 
index $\ell$ from a gauge vector that is singular at 
the position of the particle, and this is discussed below.     
The mode-sum renormalization of the metric is described, 
for example, in Sec.~IV of \cite{detweiler08} and we briefly review 
its relevant features. (It is based on the Barack-Ori mode-sum version of the 
MiSaTaQuWa renormalization \cite{MinoSasaki,quinnwald} that is reviewed 
in detail in \cite{barack09}.)     
The components $h^{\rm ret, Lor}_{\mu\nu}$ of the perturbed 
retarded metric along an orthonormal frame have the singular 
behavior of a Coulomb field, proportional to $\rho^{-1}$, where $\rho$ is the 
geodesic distance to the particle trajectory. This implies that  
$H^{\rm ret, Lor}$ has the same Coulomb singular behavior.  The angular harmonics 
$H^{\rm ret, Lor}_{\ell m} Y_{\ell m}$ of $H^{\rm ret, Lor}$ then have finite 
limits $H_{\ell m} = \lim_{r\rightarrow r_0} H^{\rm ret, Lor}_{\ell m}(t=0,r)$ 
on a sphere through the trajectory of the particle.  At the position of the 
particle, the projection 
\be
  H^{\rm ret, Lor}_\ell = \sum_m H_{\ell m} Y_{\ell m}(\pi/2,0) 
\ee
onto the $\ell$th subspace has the form  
\be 
  H^{\rm ret, Lor}_\ell = H^{\rm S, Lor}_\ell + H^{\rm ren}_\ell, 
\ee    
where $H^{\rm ren}_\ell$, the value of the renormalized field, falls off faster than 
any power of $\ell$.  The singular field $H^{\rm S, Lor}$ has $\ell$-dependence  
\be
  H^{\rm S, Lor}_\ell = E_0 L^0 + O(L^{-2}), \qquad 
\label{eq:hs}\ee
with $E_0$ independent of $L$; and the sum over $\ell$ of the $O(L^{-2})$ terms 
vanishes.  That is, $H^{\rm S} = H^{\rm s} + O(\rho)$, where $H^{\rm S}_\ell = E_0$.
The value of $H^{\rm ren}$ is then given by 
\be
 \lim_{\ell_{\rm max}\rightarrow \infty}\sum_{\ell=0}^{\ell_{\rm max}} H^{\rm ren}_\ell = \lim_{\ell_{\rm max}\rightarrow \infty}\sum_{\ell=0}^{\ell_{\rm max}} (H^{\rm ret}_\ell - H^{\rm s}_\ell).
\label{eq:hren}\ee

In our radiation gauge, we will find that, although $H^{\rm S, ORG}_\ell$ again has the form
\be
  H^{\rm S, ORG}_\ell = E_0 L^0 + O(L^{-2}), 
\label{eq:hsr}\ee
with $E_0$ agreeing with its value for a Lorenz gauge, the $O(L^{-2})$ part gives 
a contribution of order $\rho^0$ that is odd under parity in a hypersurface orthogonal 
to the trajectory: In particular, the sum of the $O(\rho^0)$ contributions from the 
limits $r\rightarrow r_0^+$ and $r\rightarrow r_0^-$ vanishes.  

The loss of gauge invariance at subleading order follows from the 
way one proves invariance of $H^{\rm ren}$ for a helically symmetric gauge vector 
$\xi^\alpha$ that is differentiable at the position of the particle (implying, 
in particular, $H^{\rm ren,Lor} = H^{\rm ren,ORG}$).
Under a gauge transformation with $u^\alpha$ fixed, $H$ changes by 
$\frac12\Lie_{\bm\xi} g_{\alpha\beta}u^\alpha u^\beta = \na_\alpha\xi_\beta u^\alpha u^\beta$.
From Eqs.~(\ref{eq:ut}) and (\ref{eq:k}) for $u^\alpha$, we have   
\be
  \na_\alpha\xi_\beta u^\alpha u^\beta = u^t u^\beta k^\alpha\na_\alpha\xi_\beta 
	= u^t u^\beta(\Lie_{\bm k} \xi_\beta - \xi_\alpha\na_\beta k^\alpha) = - \xi_\alpha u^\beta\na_\beta u^\alpha 
= 0,  
\label{gaugeinv}\ee  
where we have used the geodesic equation in the last equality.  
Now $u^\alpha$, as defined 
by Eq.~(\ref{eq:ut}), satisfies the geodesic equation only on the particle's trajectory; 
for points a geodesic distance $\rho$ from the trajectory, it satisfies 
\be
   u^\beta\nabla_\beta u^\alpha = O(\rho).
\ee
The gauge vector $\xi^\alpha$ relating a Lorenz gauge to our radiation gauge diverges as 
$\log\rho$ near $\rho=0$ \cite{bo01}, leading to a term of order $\rho^0$ in 
Eq. (35).  We defer to Sec.~\ref{sec5}, in which the explicit analytic construction of 
the perturbed metric is presented, the discussion of the parity of this term.  

In the actual computation, as described in Sec.~(\ref{sec7}), we use spin-weighted 
spherical harmonics instead of ordinary spherical harmonics for parts of the metric 
with different spin weights, and we then check 
that the resulting leading term in the singular field -- the value of $E_0$ -- 
agrees with its value computed analytically within a Lorenz-gauge framework 
by Linz \cite{linz}.

\subsection{Numerical methods - Angular harmonics}
\label{sec2c}

This section describes two different series expressions for spin-weighted spheroidal harmonics 
that we use for two different purposes:  
First, following Fackerell and Crossman \cite{FackerellCrossman}, we write each spin-weighted 
spheroidal harmonic as a sum of Jacobi polynomials. The formalism provides an accurate 
way to evaluate the spheroidal harmonics and the angular eigenvalues on 
which both the angular harmonics and the radial functions $R_{\ell m}$ depend.  Second, 
to renormalize the metric and the self-force, we write each spin-weighted spheroidal harmonic as sum of spin-weighted {\rm spherical} harmonics.  Although the present paper uses only harmonics 
$\dis{\,}_sS_{\ell m } := {\,}_sS_{\ell m \omega}|_{\omega=m\Omega}$, the formalism 
is developed for ${\,}_sS_{\ell m \omega}$, with no restriction on $\omega$.   

For $\gamma = a\omega$, the spin-weighted spheroidal harmonics satisfy the eigenvalue equation
\begin{align}\label{Sequation}
& \frac{1}{\sin\theta}\frac{d}{d\theta}\left(\sin\theta\frac{d}{d\theta}\,{\,}_sS_{\ell m \omega}\right) + \Biggl[ \gamma^2\cos^2\theta-2s\gamma\cos\theta - \frac{2ms\cos\theta+s^2+m^2}{\sin^2\theta}+{\,}_sE_{\ell m \omega} \Biggr]{\,}_sS_{\ell m \omega} = 0, 
\end{align}
for a Teukolsky-equation source with time-dependence 
$e^{i\omega t}$.  The eigenvalue ${\,}_sE_{\ell m \omega}$ is a continuous function of 
$\gamma$ that takes the value $\ell(\ell+1)$ when $\gamma = 0$. (Note that the 
spin-weighted spheroidal harmonics and their eigenvalues depend 
on $\omega$ only via $\gamma$, so the conventional use of the index 
$\omega$ instead of $\gamma$ is slightly misleading.)

For fixed $s$ and $\gamma$, with $\gamma$ real, the eigenfunctions are complete and orthogonal on the sphere, satisfying
\bea
\delta(\cos\theta-\cos\theta^\prime)\delta(\phi-\phi^\prime) &=& \sum_{\ell m}{\,}_sS_{\ell m \omega}(\theta)e^{im\phi}{\,}_sS_{\ell m \omega}(\theta^\prime)e^{-im\phi^\prime} ,
\eea
\bea
\int_0^\pi \int_0^{2\pi} {\,}_sS_{\ell m \omega}(\theta)e^{im\phi} {\,}_sS_{\ell^\prime m^\prime\omega}(\theta)e^{im'\phi} \sin\theta d\theta d\phi&=& \delta_{\ell,\ell^\prime}  \delta_{m,m^\prime}.
\eea

When $\gamma = 0$, ${\,}_sS_{\ell m\omega}(\theta)e^{im\phi}$ becomes the spin-weighted spherical harmonic ${}_sY_{\ell m}(\theta,\phi)$, given by
\be
{}_sY_{\ell m}= \left\{\begin{array}{ll}
              \left[ (\ell-s)!/(\ell+s)! \right]^{1/2}\eth^s Y_{\ell m}, &\ \ 0\le s\le \ell,
              \\
              (-1)^s\left[(\ell+s)!/(\ell-s)!\right]^{1/2}\bar{\eth}^{-s} Y_{\ell m},&
         -\ell\le s\le 0,
        \end{array}\right.
 \label{green_sylm_a}\ee
with 
\begin{eqnarray}
\eth\eta &=& -\left(\partial_\theta+i\csc\theta\partial_\phi-s\cot\theta\right)\eta, 
\label{green_eth_b}
\nonumber\\
\bar{\eth}\eta&=&-\left(\partial_\theta-i\csc\theta\partial_\phi+s\cot\theta\right)\eta
\label{green_eth_bar_b}
\end{eqnarray}
where $\eth$ and $\bar{\eth}$ are, respectively, raising and lowering operators for the spin-weight, and $\eta$ is a quantity of spin-weight $s$.\\

\subsubsection{Spin-weighted spheroidal harmonics as a sum over Jacobi Polynomials}
\label{JP}

The formalism that expresses the angular harmonic ${}_sS_{\ell m \omega}$ in 
terms of Jacobi polynomials involves several constants that depend on parameters
$s,\ell,m$, and $\gamma$ whose values are fixed in 
this section.  To avoid encumbering a large number of symbols with the four indices $s,m,\ell,\omega$, 
in this and the next subsections (Sects. \ref{JP} and \ref{sYlm}) we suppress the indices, so that, for example the angular harmonic and its corresponding eigenvalue will be written as 
\be
      S\equiv {}_sS_{\ell m \omega}, \qquad E:= {}_sE_{\ell m\omega}.
\ee

To calculate the spin-weighted spheroidal harmonics, their derivatives and eigenvalue, we set $x:=\cos\theta$ and write the homogenous angular equation, Eq. (\ref{Sequation}), as follows:
\bea
(1-x^2)\frac{d^2S}{dx^2} - 2x\frac{dS}{dx} + \left(\gamma^2 x^2 -2\gamma s x - \frac{m^2+s^2+2msx}{1-x^2} + E \right)S = 0,
\eea
The eigenfunctions' dominant behavior at $x=\pm1$ is $(1\mp x)^{|m\pm s|/2}$. Following the 
formalism (and notation) of Fackerell and Crossman \cite{FackerellCrossman}, we introduce $\alpha = |m+s|$ and $\beta = |m-s|$ for simplicity and introduce new functions $U$ and $V$, as follows:
\bea \label{SUV1}
S (x) &=& e^{\gamma x}\left(\frac{1-x}{2}\right)^\alpha \left(\frac{1+x}{2}\right)^\beta U(x) {\,}, \\ \label{SUV2}
S (x) &=& e^{-\gamma x}\left(\frac{1-x}{2}\right)^\alpha \left(\frac{1+x}{2}\right)^\beta V(x).
\eea
The functions $U$ and $V$ satisfy the following differential equation (where the upper sign is used 
if $F=U$, the lower sign if $F=V$):
\bea
&&(1-x^2)\partial_x^2F+[\beta-\alpha-x(2+\alpha+\beta)\pm2\gamma(1-x^2)]\partial_xF \nonumber \\
&& + \Biggl[ E+\gamma^2-\frac{\alpha+\beta}{2}\left( \frac{\alpha+\beta}{2}+1\right)\pm\gamma(\beta-\alpha)\mp\gamma x(\alpha+\beta+2\pm2s)\Biggr]F = 0.
\eea
The above differential equation is closely related to that of the Jacobi polynomial given by the Rodgrigues formula
\bea
P_n^{(\alpha,\beta)}(x) = \frac{(-1)^n}{2^n n!}(1-x)^{-\alpha}(1+x)^{-\beta}\partial_x^n\left[ (1-x)^{\alpha_+n}(1+x)^{\beta+n}\right],
\eea
which satisfies
\bea
\left[(1-x^2)\partial_x^2+[\beta-\alpha-x(\alpha+\beta+2)]\partial_x+n(n+1+\alpha+\beta)\right]P_n^{(\alpha,\beta)}=0.
\eea
Expanding the functions $U$ and $V$ as infinite series of Jacobi polynomials,
\bea \label{UV}
U (x) &=& \sum_{r=0}^\infty A^{(r)}P_r^{(\alpha,\beta)}(x), \nonumber \\
V (x) &=& \sum_{r=0}^\infty B^{(r)}P_r^{(\alpha,\beta)}(x), 
\eea
and using the recurrence relations satisfied by the Jacobi polynomials, we get the following two recurrence relations for $A^{(r)}$ and $B^{(r)}$ (where the upper sign is used if $G = A$ and the lower if $G = B$):
\begin{align}\label{ola1}
\left[ E + \gamma^2 - \frac{\alpha+\beta}{2}\left( \frac{\alpha+\beta}{2}+1\right)+\frac{2s\gamma(\alpha-\beta)}{\alpha+\beta+2}\right]G^{(0)} \pm \left[\frac{4\gamma(\alpha+1)(\beta+1)\left( \frac{\alpha+\beta}{2}+1\mp s\right)}{(\alpha+\beta+2)(\alpha+\beta+3)}\right]G^{(1)} = 0,
\end{align}
\begin{align} \label{ola2}
& \left[\pm \frac{4\gamma(r+\alpha+1)(r+\beta+1)\left(r+1 \mp s+\frac{\alpha+\beta}{2} \right)}{(2r+\alpha+\beta+2)(2r+\alpha+\beta+3)}\right] G^{(r+1)} \mp \left[ \frac{4\gamma r(r+\alpha+\beta)\left( r\pm s+\frac{\alpha+\beta}{2}\right)}{(2r+\alpha+\beta-1)(2r+\alpha+\beta)}\right] G^{(r-1)} \nonumber \\
&\hskip2.2cm + \Biggl[ E+\gamma^2-\left( r+\frac{\alpha+\beta}{2}\right)\left( r+\frac{\alpha+\beta}{2}+1\right) + \frac{2\gamma s(\alpha-\beta)(\alpha+\beta)}{(2r+\alpha+\beta)(2r+\alpha+\beta+2)}\Biggr]G^{(r)} = 0.
\end{align} 
\\ \\
\emph{Determination of ${\,}_sE_{\ell m}^\gamma$}\\

For the series in Jacobi polynomial to converge, the constant $E$ should satisfy a transcendental equation. To enforce this, we define the following quantities:
\bea
N^{(r)} &=& \frac{2\gamma(r+\alpha)(r+\beta)(2r+\alpha+\beta-2s)}{(2r+\alpha+\beta)(2r+\alpha+\beta+1)}\frac{A^{(r)}}{A^{(r-1)}} \\
K^{(r)} &=& \left(r+\frac{\alpha+\beta}{2} \right)\left(r+\frac{\alpha+\beta}{2}+1 \right)-\gamma^2-\frac{2\gamma s(\alpha-\beta)(\alpha+\beta)}{(2r+\alpha+\beta)(2r+\alpha+\beta+2)} \\
L^{(r)} &=& \frac{4r\gamma^2(r+\alpha)(r+\beta)(r+\alpha+\beta)(2r+\alpha+\beta+2s)(2r+\alpha+\beta-2s)}{(2r+\alpha+\beta)^2(2r+\alpha+\beta+1)(2r+\alpha+\beta-1)}
\eea
where none of the $A^{(r)}$s vanish. Eqs. ($\ref{ola1}$) and ($\ref{ola2}$) may be written as 
\bea \label{ola11}
N^{(1)}-K^{(0)}+E=0 
\eea
and
\bea \label{ola22}
N^{(r+1)} = K^{(r)} - E + \frac{L^{(r)}}{N^{(r)}}
\eea
which by iterating gives us
\bea\label{ola222}
N^{(r+1)} = K^{(r)}  - E + \frac{L^{(r)}}{K^{(r-1)}-E+\frac{L^{(r-1)}}{K^{(r-2)}-E+\frac{L^{(r-2)}}{K^{(r-3)}-E+\frac{L^{(r-3)}}{K^{(r-4)}-E+\cdots+\frac{L^{(1)}}{K^{(0)}-E}}}}}
\eea
Equation ($\ref{ola22}$) can also be written as
\bea
N^{(r)} = \frac{L^{(r)}}{E-K^{(r)}+N^{(r+1)}}
\eea
which would then give us 
\bea\label{ola33}
N^{(r+1)} = -\frac{L^{(r+1)}}{K^{(r+1)}-E+\frac{L^{(r+2)}}{K^{(r+2)}-E+\frac{L^{(r+3)}}{K^{(r+3)}-E+\frac{L^{(r+4)}}{K^{(r+4)}-E+\cdots}}}}.
\eea
Equating Eqs. ($\ref{ola222}$) and ($\ref{ola33}$), we get a transcendental equation for $E$, namely
\begin{align}
E=&K^{(r)} +\frac{L^{(r)}}{K^{(r-1)}-E+\frac{L^{(r-1)}}{K^{(r-2)}-E+\frac{L^{(r-2)}}{K^{(r-3)}-E+\cdots+\frac{L^{(1)}}{K^{(0)}-E}}}} \\ \nonumber
&+ \frac{L^{(r+1)}}{K^{(r+1)}-E+\frac{L^{(r+2)}}{K^{(r+2)}-E+\frac{L^{(r+3)}}{K^{(r+3)}-E+\cdots}}}.
\end{align}
We solve this in Mathematica to machine precision.  
Convergence of the infinite continued fraction in Eq. ($\ref{ola33}$) is treated in the references mentioned in \cite{FackerellCrossman}.
\\ \\
\emph{Normalization of $S(x)$ }

We have already expanded the $S(x)$ as a sum over Jacobi polynomials and have the recurrence relations for the coefficients in the expansion where all the coefficients depend on either $B^{(0)}$ or $A^{(0)}$ in Eq. ($\ref{UV}$). Now we have,
\bea \label{SintermsofP}
S(x) = e^{\gamma x}\left( \frac{1-x}{2}\right)^{\alpha/2}\left( \frac{1+x}{2}\right)^{\beta/2}\sum_{r=0}^\infty A^{(r)}P_r^{(\alpha,\beta)}(x) \nonumber \\
S(x) = e^{-\gamma x}\left( \frac{1-x}{2}\right)^{\alpha/2}\left( \frac{1+x}{2}\right)^{\beta/2}\sum_{r=0}^\infty B^{(r)}P_r^{(\alpha,\beta)}(x).
\eea
For the eigenfunctions to be normalized, we need two equations for the two unknowns, $A^{(0)}$ and $B^{(0)}$. We get one from the Eqs. (\ref{SUV1}) and (\ref{SUV2}) which gives us
\bea
V = e^{2\gamma x} U.
\eea
Using Eq. (\ref{UV}) at $x$ = 1, we have
\bea \label{unknown1}
\frac{B^{(0)}}{A^{(0)}} = e^{2\gamma}\left( \sum_{r=0}^\infty \frac{A^{(r)}}{A^{(0)}}\frac{(r+\alpha)!}{r!\alpha!} \right)/\left(  \sum_{r=0}^\infty \frac{B^{(r)}}{B^{(0)}}\frac{(r+\alpha)!}{r!\alpha!} \right).
\eea
The other equation is obtained by using 
\bea
\int_{-1}^1 S^2dx = \frac{1}{2\pi}
\eea
We use $(2\pi)^{-1}$ here so that $S(x)e^{im\phi}$ is normalized to 1. Using this normalization condition along with Eq. (\ref{SintermsofP}), we have
\bea
A^{(0)}B^{(0)}\sum_{n,r=0}^\infty  \frac{A^{(n)}}{A^{(0)}}\frac{B^{(r)}}{B^{(0)}}\int_{-1}^1\left( \frac{1-x}{2}\right)^\alpha\left( \frac{1+x}{2} \right)^\beta P_n^{(\alpha,\beta)}(x) P_r^{(\alpha,\beta)}(x)dx = \frac{1}{2\pi},
\eea
which gives us
\bea
A^{(0)}B^{(0)}\sum_{n,r=0}^\infty  \frac{A^{(n)}}{A^{(0)}}\frac{B^{(r)}}{B^{(0)}}\frac{2}{2n+\alpha+\beta+1}\frac{\Gamma(n+\alpha+1)\Gamma(n+\beta+1)}{n!\Gamma(n+\alpha+\beta+1)}\delta_{n,r} = \frac{1}{2\pi}.
\eea
Hence, the second equation is 
\bea\label{unknown2}
\frac{1}{A^{(0)}B^{(0)}} = 2\pi\sum_{n=0}^\infty  \frac{A^{(n)}}{A^{(0)}}\frac{B^{(n)}}{B^{(0)}}\frac{2}{2n+\alpha+\beta+1}\frac{\Gamma(n+\alpha+1)\Gamma(n+\beta+1)}{n!\Gamma(n+\alpha+\beta+1)}.
\eea
The ratios $A^{(r)}/A^{(0)}$ and $B^{(r)}/B^{(0)}$ are easily calculated by using their recurrence relations Eqs. (\ref{ola1}) and (\ref{ola2}) and using $A^{(r)}$ = $B^{(r)}$ = 0 for $r<0$. Therefore Eqs. (\ref{unknown1}) and (\ref{unknown2}) correctly determine the first coefficients in the expansion by choosing the $A^{(0)}$ whose real part is positive.

\subsubsection{Spin-weighted spheroidal harmonics as a sum over ${\,}_sY_{\ell m}$} 
\label{sYlm}

The spectral decomposition of spin-weighted spheroidal harmonics in terms of spin-weighted spherical harmonics as in \cite{PressTeukolsky}, \cite{hughes00} has the form
\bea
S (\theta) e^{im\phi}= \sum_{j= \ell_{\rm min}}^\infty b_j{\,}_sY_{j,m}(\theta,\phi),
\label{eq:SbY}\eea
where $\ell_{\rm min}  = \rm{max}(|m|,|s|)$. Substituting the above in Eq. (\ref{Sequation}), we have
\bea \label{eqn59}
\sum_j b_j\left[ \gamma^2\cos^2\theta-2\gamma s \cos\theta-j(j+1)\right]{\,}_sY_{j,m} = -E \sum_j b_j{\,}_sY_{j,m}.
\eea
We now write $\cos^n\theta$ for $n=0,1,2$ as sums of spherical harmonics. It is then easy to group them and the spin-weighted spherical harmonics as follows: From the relations
\bea
\cos\theta = 2\sqrt{\frac{\pi}{3}} Y_{1,0}{\,}, \quad \cos^2\theta = \frac{4\sqrt{\pi}}{3\sqrt{5}}Y_{2,0} + \frac{1}{3} {\,},
\eea
\bea \label{comine}
D^{j_1}_{\mu_1m_1}D^{j_2}_{\mu_2m_2} = \sum_{j,\mu,m} \langle j_1,\mu_1; j_2,\mu_2|j,\mu\rangle\langle j_1,m_1;j_2,m_2|j,m\rangle D^j_{\mu,m} {\,},
\eea
and
\bea
{\,}_sY_{\ell m}= \sqrt{\frac{2\ell+1}{4\pi}}D^\ell_{-s,m},
\eea
we find (for $n=0,1,2$)
\be\label{niceCGeqn}
\cos^n\theta {\,}_sY_{\ell m} = \frac{n}{2n-1}\sum_{j=\ell-n}^{\ell+n} \sqrt{\frac{2\ell+1}{2j+1}}\langle\ell,-s;n,0|j,-s\rangle\langle\ell,m;n,0|j,m\rangle{\,}_sY_{j,m} + \frac{\delta_{n,2}}{3}{\,}_sY_{\ell m} + \delta_{n,0}{\,}_sY_{\ell m}.
\ee
Multiplying  Eq (\ref{eqn59}) with ${\,}_sY_{k,m}$, integrating over the 2-sphere, and using Eq (\ref{niceCGeqn})
we have
\bea\label{Srecurrence}
&& b_{k-2}\left[\gamma^2c_{k-2,k,2}\right] + b_{k-1} \left[ \gamma^2c_{k-1,k,2}-2s\gamma c_{k-1,k,1}\right] + b_{k}\left[ \gamma^2 c_{k,k,2} - 2s\gamma c_{k,k,1}-k(k+1)\right] \nonumber \\
&& + b_{k+1}\left[ \gamma^2c_{k+1,k,2}-2s\gamma c_{k+1,k,1}\right] + b_{k+2}\left[ \gamma^2c_{k+2,k,2}\right] = -E b_{k},
\eea
where 
\bea
c_{k,j,2} = \frac{\delta_{k,j}}{3} + \frac{2}{3}\sqrt{\frac{2j+1}{2k+1}} \langle j,m;2,0|k,m \rangle \langle j,-s;2,0|k,-s\rangle \nonumber \\
c_{k,j,1} = \sqrt{\frac{2j+1}{2k+1}}\langle j,m;1,0|k,m\rangle \langle j,-s;1,0|k,-s\rangle.
\eea
Eq. (\ref{Srecurrence}) can be written as a matrix equation where the $b$ are the matrix's eigenvector and $E$ are the eigenvalues. It is then easy to solve the matrix equation for the eigenvectors and eigenvalues as the matrix is band diagonal. Here $\langle j_1,m_1;j_2,m_2|j,m\rangle$ are the Clebsch-Gordan coefficients.

To calculate ${\,}_sY_{\ell m}$ to high precision, we used the following analytical forms of 
spin-weighted harmonics at $\theta=\pi/2$.  Introducing the symbol  
$\displaystyle 
         e_{\ell m} := \begin{cases} 1 & \ell+m\ \mbox{ even}\\
                0, & \ell+m\ \mbox{ odd}\ ,
\end{cases}
$
we can write
\bea
Y_{\ell m}(\frac{\pi}{2},0) &=&  (-1)^{(\ell+m)/2} \sqrt{\frac{(2\ell+1)}{4\pi}}
			\frac{\sqrt{ (\ell-m)!(\ell+m)!}}{\quad(\ell-m)!!(\ell+m)!!}\ e_{\ell m}\ \,
\label{ypi2}\\  
\nonumber\\
{\,}_1Y_{\ell m}(\frac{\pi}{2},0) 
&=& (-1)^{(\ell+m)/2}\sqrt{\frac{(2\ell+1)(\ell-m)!(\ell+m)!}{4\pi\ell(\ell+1)}} \left[  \frac{m\,e_{\ell,m}}{(\ell-m)!!(\ell+m)!!} - \frac{i\,e_{\ell,m+1}}{(\ell-m-1)!!(\ell+m-1)!!} \right],
%2^{-\ell}\sqrt{\frac{(2\ell+1)}{4\pi \ell(\ell+1)(\ell-m)!(\ell+m)!}} 
   %     \left[ (-1)^{(\ell+m)/2} m e_{\ell m} - (-1)^{(\ell+m+1)/2}(\ell-m)(\ell+m) e_{\ell,m+1}\right],
\label{1ypi2}\\ 
{\,}_2Y_{\ell m}(\frac{\pi}{2},0) 
&=& (-1)^{(\ell+m)/2} \sqrt{\frac{(2\ell+1)(\ell-m)!(\ell+m)!}{4\pi(\ell-1)\ell(\ell+1)(\ell+2)}} \left[  \frac{[2m^2-\ell(\ell+1)]\,e_{\ell,m}}{(\ell-m)!!(\ell+m)!!} - \frac{2im\,e_{\ell,m+1}}{(\ell-m-1)!!(\ell+m-1)!!} \right].
%2^{-\ell}\sqrt{\frac{(2\ell+1)}{4\pi \ell(\ell+1)(\ell-m)!(\ell+m)!}} \times\nonumber\\
  %      && \phantom{xxx} \left\{ (-1)^{(\ell+m)/2}[2m^2-\ell(\ell+1)]e_{\ell m} 
    %            - (-1)^{(\ell+m+1)/2}2m(\ell-m)(\ell+m)e_{\ell,m+1}\right\}.
\label{2ypi2}\eea
Eq.~(\ref{ypi2}) is quickly obtained from the corresponding equation for 
$P_\ell^m(0)$, given, for example, in Arfken and Weber \cite{arfken}; 
the corresponding 
relations (\ref{1ypi2}) and (\ref{2ypi2}) for spin-weighted harmonics follow from their 
definition (\ref{green_sylm_a}), with recurrence relations of associated Legendre 
polynomials used to eliminate $\theta$ derivatives.

Values of the angular harmonics $\,_2S_{\ell m\omega}$ at $(\theta,\phi)= (\pi/2,0)$, 
computed as a sum of Jacobi polynomials and as a sum over spin-weighted spherical harmonics 
$\,_2Y_{\ell m}$  are listed in Table~I to show the accuracy of our calculation 
of these angular eigenfunctions.

\begin{center}
\begin{table}[h]
  \begin{tabular}{|c|c|c|c|c|c|}
    \hline
                 \multicolumn{1}{|p{0.7cm}|}
                 {\centering $\ell$}
                & \multicolumn{1}{|p{0.7cm}|}
                 {\centering $m$}
                & \multicolumn{1}{|p{1.0cm}|}
                 {\centering $a/M$}
                 & \multicolumn{1}{|p{1.0cm}|}
                 {\centering $r_0/M$}
                 & \multicolumn{1}{|p{4.2cm}|}
                 {\centering $\,_sS_{\ell,m}$ ($P_r^{(\alpha,\beta)}$)}
                 & \multicolumn{1}{|p{4.2cm}|}
                 {\centering $\,_sS_{\ell,m}$ ($\,_sY_{\ell,m}$)}  \\

    \hline
       6        &        3        &        0.90                &        2.321        &        0.25240458701173892108        &        0.25240458701173889664        \\
               10        &        8        &        -0.80        &        8.432        &        -0.077523625602031470364        &        -0.077523625602031470355        \\
        15        &        -14        &        0.56                &        3.994        &        -0.39964402300714286677        &        -0.39964402300714286644        \\
        20        &        20        &        0.95                &        1.938        &        0.53866543681715165119        &        0.53866543680910100221        \\
        25        &        -24        &        0.75                &        3.159        &        -0.38677474564628361398        &        -0.38677474564628311527        \\
        30        &        1        &        0.69                &        3.439        &        0.018864751317113632621        &        0.018864751317113632621        \\
        35        &        -29        &        0.43                &        4.502        &        -0.12392842343512166756        &        -0.12392842343512166731        \\
        41        &        38        &        0.85                &        2.633        &        0.41329611968515525721        &        0.41329611968514376716        \\
        45        &        43        &        -0.42        &        7.315        &        -0.44047075216769495176        &        -0.44047075216769495195        \\
        50        &        -47        &        0.50                &        4.234        &        0.37391460514301075256        &        0.37391460514301011670        \\
        54        &        43        &        0.29                &        5.015        &        0.39632284051223687540        &        0.39632284051223687540        \\
        60        &        58        &        0.81                &        2.860        &        -0.46730374640820672293        &        -0.46730374640803870557        \\
        65        &        42        &        -0.40        &        7.255        &        -0.35895613544467811490        &        -0.35895613544467811490        \\
        70        &        -70        &        0.67                &        3.529        &        -0.83907652666117403240        &        -0.83907652666117381586        \\
        75        &        74        &        0.80                &        2.910        &        0.32010913665973882714        &        0.32010913665976780184        \\        
        80        &        78        &        0.64                &        3.660        &        -0.54551806835154401208        &        -0.54551806835154351539        \\
        85        &        -85        &        0.55                &        5.555        &        -0.88887357826499020752        &        -0.88887357826499020734        \\
                                                
    \hline
  \end{tabular}
\caption{For each listed value of $\ell$, $m$, $a$ and $r_0$, we give the value of $\,_2S_{\ell,m}^\omega$, obtained by using the formalism given in subsections (\ref{JP}) and (\ref{sYlm}), with 
$\gamma = a\,m\,\Omega = \frac{a\,m\,M^{1/2}}{r_0^{3/2}+a}$ and $\theta=\pi/2$. The fractional 
accuracy increases with increasing $r_0$, and all except the last five values of 
$r_0$ are chosen to be within a few percent of the innermost stable circular orbit for a 
given $a$.}
  \end{table}\label{tableI}
\end{center}

\subsection{Numerical methods -  radial harmonics} 
\label{sec2d}
\subsubsection{Teukolsky equation}

We integrate $R_H$ and $R_\infty$ from the horizon and infinity, respectively. The homogenous solutions $R_H$ and $R_\infty$ at the horizon and infinity are given by the following series
\bea
R_H &=& \frac{e^{-i\omega r_\star}}{\Delta^2}\sum_{n=0}^\infty c_n\left(\frac{r-r_+}{M}\right)^n,\\
R_\infty &=& e^{i\omega r_\star}\sum_{n=0}^\infty \frac{d_n}{(r/M)^{n+5}},  
\eea
where 
\bea
r_\star &=& r + \frac{r_+^2+a^2}{r_+-r_-}\ln|r-r_+| - \frac{r_-^2+a^2}{r_+-r_-}\ln|r-r_-| \quad {\rm and}\\
r_{\pm} &=& M \pm \sqrt{M^2-a^2}.
\eea
The expansion coefficients satisfy the following recurrence relations
\begin{align}
c_n =& \Biggl\{2i(n-6)\omega M c_{n-3} + \biggl[E-n^2+7n-12 +\omega^2 a^2 
        + i(8n-34)\omega M \sqrt{1-a^2/M^2} + i (4n-14) \omega M \biggr] c_{n-2}\nonumber \\ 
   & \quad + \biggl[4 m \omega a+(2E -4 n^2 +18 n -18 +2\omega^2 a^2)\sqrt{1-a^2/M^2} \nonumber \\
 & \qquad\qquad + i 12(n-2)\omega M -i 4(2n-5)\omega a^2/M -i 4ma/M +i 12(n-2)\omega M\sqrt{1-a^2/M^2} \biggr] c_{n-1} \Biggr\}\nonumber \\   
%%\displaybreak
 &\hskip4mm \Biggl\{ (4n^2-8n)(1-a^2/M^2) + m^2 a^2/M^2 -4m\omega a -4m\omega a\sqrt{1-a^2/M^2} \nonumber\\
 &\hskip5cm -i[ 8n M\omega (1-a^2/M^2) +(8n\omega M-4ma/M)\sqrt{1-a^2/M^2}\Biggr\}^{-1},\\
d_n = &-\frac{i(n+1)a^4}{2\omega M^5}d_{n-5} + \frac{i(n+1)(2n+1+i\omega a^2/M)a^2}{n\omega M^3}d_{n-4} \nonumber \\
& + \frac{- 4mMa + (4n+8)\omega M a^2 - i(4n^2+8n)M^2 +i(E-2n^2-4n-m^2)a^2 +i\omega^2 a^4}
        {2n\omega M^2} \ d_{n-3} \nonumber \\ 
& + \frac{2ma-(2n+1)\omega a^2 + i(-E+2n^2+5n+3 + 2m\omega a -\omega^2  a^2)M}{n\omega M}d_{n-2} \nonumber \\
&+ \frac{(4n-4)\omega M +i(E-n^2-3n-2+\omega^2 a^2)}{2n\omega}d_{n-1}.
\end{align}
We use a 7th order Runge-Kutta routine to solve for the homogenous radial solutions using the above initial/boundary conditions, obtaining values of $\psi_0^\textrm{ret}$ to an accuracy of 1 part in $10^{13}$.
 
\subsubsection{Sasaki-Nakamura equation}
The Sasaki-Nakamura equation is
\bea
\frac{d^2X}{dr^{*2}} - F\frac{dX}{dr^*} - UX=0
\eea
where $X$ is related to the radial part, $R_4$ of $\rho^{-4}\psi_4$ by
\bea
R_4 = \frac{1}{\eta} \left[\left(\alpha + \frac{\beta^\prime}{\Delta} \right)\chi - \frac{\beta}{\Delta}\chi^\prime\right]
\eea
where
\bea
\chi = \frac{\Delta X}{\sqrt{r^2+a^2}}.
\eea
 The radial parts, $R_4$ and $R_0$ are related to each other by the relation
 \bea
 R_0 = c \frac{R_4^*}{\Delta^2},
 \eea
 where $c$ is a constant.
 The function $F$ is
 \begin{align}
 F &= \frac{\eta^\prime}{\eta} \frac{\Delta}{r^2+a^2}
 \end{align}
 where
 \begin{align}
 \eta = &-12i\omega M + \lambda(\lambda+2) - 12a\omega(a\omega-m)+ \frac{8ia[3a\omega - \lambda(a\omega - m)]}{r} \nonumber \\
 &+ \frac{-24iaM(a\omega-m) + 12a^2[1-2(a\omega-m)^2] }{r^2} + \frac{24ia^3(a\omega-m) - 24Ma^2 }{r^3} + \frac{12a^4}{r^4}
 \end{align}
 and the prime denotes a derivate with respect to $r$.
 \\The function $U$ is
 \begin{align}
 U &= \frac{\Delta U_1}{(r^2+a^2)^2} + G^2 + \frac{\Delta G^\prime}{r^2+a^2} - F G
 \end{align}
 where $U_1$ and $G$ are
 \begin{align}
 U_1 &= V + \frac{\Delta^2}{\beta} \left[ \frac{d}{dr}\left( 2\alpha + \frac{\beta^\prime}{\Delta} \right) - \frac{\eta^\prime}{\eta} \left( \alpha + \frac{\beta^\prime}{\Delta}\right) \right] \, {\rm and}\nonumber \\
 G &= - \left[ \frac{2(r-M)}{r^2+a^2} \right] + \frac{r\Delta}{(r^2+a^2)^2}.
 \end{align}
 The $\alpha$, $\beta$ and $V$ that appear above are given by
  \begin{align}
V &= -\left[ \frac{ K^2 + 4iK(r-M) }{ \Delta } \right] + 8i\omega r+ \lambda\, , \nonumber \\
\beta &= 2\Delta\left[ -iK+r-M-\frac{2\Delta}{r} \right]\, , \nonumber \\
\alpha &= \frac{-iK\beta}{\Delta^2} + 3iK^\prime + \lambda + \frac{6\Delta}{r^2}\, ,  \nonumber \\
\end{align}
where $K = (r^2+a^2)\omega - am$. We use the following boundary conditions at the horizon and at infinity,
\begin{align}
X_{H} &= e^{i\omega r^*} \\
X_\infty &= \sum_{n=0}^{4} e^{-i\omega r^*} \frac{\tilde{d}_n}{r^n}\,\,
\end{align}
where
\begin{align}
r^* &= r + \frac{2Mr_+}{r_+-r_-} \log \left|\frac{r-r_+}{2M} \right| - \frac{2Mr_-}{r_+-r_-} \log \left|\frac{r-r_-}{2M} \right| 
\end{align}
and \cite{hughes00}
\begin{align}
\tilde{d}_0 &= 1 \nonumber \\
\tilde{d}_1 &= \frac{-i(2+\lambda+2am\omega)}{2\omega} \,,\nonumber \\
\tilde{d}_2 &= -\frac{\lambda^2+\lambda(4+am\omega) -12iM\omega + 4am\omega(1+am\omega+2iM\omega)}{8\omega^2} \,,\nonumber \\
\tilde{d}_3 &= \frac{i\lambda^3}{48\omega^3} + \frac{i\lambda^2(3am\omega-1)}{24\omega^3} + \frac{i\lambda(-2-2am\omega-3iM\omega+2a^2\omega^2+3a^2m^2\omega^2+6iamM\omega^2)}{12\omega^3} \nonumber \\
&\quad + \frac{i[6iM-6am-3a\omega(a+imM)+2am\omega^2(a^2-4M^2)+(am\omega)^2(am+6iM)]}{6\omega^3} \,,\nonumber \\
\tilde{d}_4 &= \frac{\lambda^4}{384\omega^4} + \frac{\lambda^3(8am\omega-12)}{384\omega^4} + \frac{\lambda^2(12-72am\omega+48iM\omega+32a^2\omega^2+24a^2m^2\omega^2+48iamM\omega^2)}{384\omega^4} \nonumber \\ 
&\quad \frac{\lambda[80(1-am\omega)+288iM\omega+128a^2\omega^2(am\omega-1)+16(am\omega)^2(2am\omega+12iM\omega-7)-256amM^2\omega^3]}{384\omega^4} \nonumber \\
&\quad + \frac{1}{24\omega^3}[30am-30iM-6a^2\omega-15\omega(am)^2+60iamM\omega+45M^2\omega-16a^3m\omega^2-2(am)^2\omega^2 \nonumber \\
&\quad -18iM(a\omega)^2-6iM(am\omega)^2+4am(M\omega)^2+8a^4m^2\omega^3+(am)^4\omega^3+24imM(a\omega)^3+12iM(am\omega)^3 \nonumber \\
&\quad -44\omega(amM\omega)^2-48iam(M\omega)^3].
\end{align}

The accuracy of our radial eigenfunctions is shown in Table II, which exhibits 
values of $|R_{\rm in}R_{\rm out}/W|$ computed by independently integrating the Teukolsky 
equation and the Sasaki-Nakamua equation.  
 
\begin{center}
\begin{table}
  \begin{tabular}{|c|c|c|c|c|c|}
    \hline
                 \multicolumn{1}{|p{0.7cm}|}
                 {\centering $\ell$}
                & \multicolumn{1}{|p{0.7cm}|}
                 {\centering $m$}
                & \multicolumn{1}{|p{1.0cm}|}
                 {\centering $a/M$}
                 & \multicolumn{1}{|p{1.0cm}|}
                 {\centering $r_0/M$}
                 & \multicolumn{1}{|p{4.2cm}|}
                 {\centering Teukolsky}
                 & \multicolumn{1}{|p{4.2cm}|}
                 {\centering Sasaki-Nakamura}  \\

    \hline
       56        &        53        &        0.50                &        4.30                &        3.32878980028437$\times10^{-2}$        &        3.32878980028426$\times10^{-2}$        \\
       85        &        84        &        0.50                &        4.25                &        2.21479032560870$\times10^{-2}$        &        2.21479032560832$\times10^{-2}$        \\
       83        &        76        &        0.30                &        5.00                &        2.66435949501300$\times10^{-2}$        &        2.664359495012799$\times10^{-2}$        \\
       42        &        1        &        0.30                &        6.00                &        5.774432559632249$\times10^{-2}$        &        5.774432559632295$\times10^{-2}$        \\ 
       6        &        -1        &        -0.25        &        6.80                &        4.40438489554819$\times10^{-1}$        &        4.40438489554828$\times10^{-1}$        \\
       56        &        37        &        0.61                &        3.79                &        2.5750354691478$\times10^{-2}$        &        2.5750354691415$\times10^{-2}$        \\
       65        &        -58        &        -0.76        &        8.33                &        6.068437461161814$\times10^{-2}$        &        6.068437461161780$\times10^{-2}$        \\
       83        &        63        &        -0.71        &        8.18                &        4.5482452073276329$\times10^{-2}$        &        4.5482452073276363$\times10^{-2}$        \\
       76        &        74        &        0.92                &        2.20                &        1.056332700247276$\times10^{-2}$        &        1.056332700247222$\times10^{-2}$        \\
       68        &        62        &        0.16                &        5.50                &        3.62341811077144$\times10^{-2}$        &        3.62341811077135$\times10^{-2}$        \\
       79        &        67        &        0.70                &        3.40                &        1.69778531677806$\times10^{-2}$        &        1.69778531677799$\times10^{-2}$        \\
       56        &        -42        &        -0.25        &        6.80                &        5.4130389893688027$\times10^{-2}$        &        5.4130389893688006$\times10^{-2}$        \\
       79        &        71        &        0.58                &        3.95                &        2.09651683213181$\times10^{-2}$        &        2.09651683213165$\times10^{-2}$        \\
       82        &        75        &        -0.2                &        6.70                &        3.77916963861862$\times10^{-2}$        &        3.77916963861852$\times10^{-2}$        \\
       54        &        35        &        0.14                &        5.55                &        4.31642221118258$\times10^{-2}$        &        4.31642221118267$\times10^{-2}$        \\
       67        &        52        &        -0.64        &        8.00                &        5.497676877828166$\times10^{-2}$        &        5.497676877828151$\times10^{-2}$        \\
       75        &        61        &        0.78                &        3.05                &        1.52296461688768494$\times10^{-2}$        &        1.52296461688768459$\times10^{-2}$        \\
       77        &        68        &        -0.14        &        6.50                &        3.85508542936437$\times10^{-2}$        &        3.85508542936429$\times10^{-2}$        \\
       65        &        53        &        -0.72        &        8.30                &        5.946127670533624$\times10^{-2}$        &        5.946127670533644$\times10^{-2}$        \\
        81        &        66        &        -0.58        &        7.80                &        4.456906403886178$\times10^{-2}$        &        4.456906403886146$\times10^{-2}$        \\
        86        &        81        &        0.47                &        4.35                &        2.20499986904938$\times10^{-2}$        &        2.20499986904907$\times10^{-2}$        \\
        53        &        -53        &        -0.23        &        6.80                &        6.057956364463913$\times10^{-2}$        &        6.057956364463899$\times10^{-2}$        \\
        70        &        65        &        0.39                &        4.70                &        2.94548366362787$\times10^{-2}$        &        2.94548366362775$\times10^{-2}$        \\
        52        &        33        &        0.04                &        6.15                &        5.053852006520149$\times10^{-2}$        &        5.053852006520189$\times10^{-2}$        \\
        44        &        44        &        -0.34        &        7.10                &        7.645731150733178$\times10^{-2}$        &        7.645731150733167$\times10^{-2}$        \\
        70        &        68        &        0.21                &        15.00        &        1.027161976019888$\times10^{-1}$        &        1.027161976019891$\times10^{-1}$        \\
        72        &        67        &        -0.66        &        43.50        &        2.9619125920183531$\times10^{-1}$        &        2.9619125920183542$\times10^{-1}$        \\
        40        &        -39        &        0.21                &        25.00        &        3.02127126128480971$\times10^{-1}$        &        3.02127126128480916$\times10^{-1}$        \\
        61        &        58        &        0.91                &        72.50        &        5.849950541617765$\times10^{-1}$        &        5.849950541617792$\times10^{-1}$        \\
        50        &        2        &        -0.88        &        15.50        &        1.4349668057921668$\times10^{-1}$        &        1.4349668057921630$\times10^{-1}$        \\
        65        &        12        &        -0.44        &        100.00        &        7.558239183584133$\times10^{-1}$        &        7.558239183584106$\times10^{-1}$        \\
                       
    \hline
  \end{tabular}
\caption{For each listed value of $\ell$, $m$, $a$ and $r_0$, we give the value of $|R_{\rm in}R_{\rm out}/W|$, obtained by integrating the Teukolsky equation and the Sasaki-Nakamura equation. As in 
Table \ref{tableI}, the fractional 
accuracy increases with increasing $r_0$, and all except the last six values of 
$r_0$ are chosen to be within a few percent of the innermost stable circular orbit for a 
given $a$.}
  \end{table}\label{table2}
\end{center}

\section{COMPUTATION OF THE PERTURBED METRIC}
\label{sec4}
We can now use Eq.~(\ref{eq:hab}) to compute the perturbed metric in an ORG in terms of the Hertz potential $\Psi$.  We first convert the spin-weighted spheroidal harmonics in Eq.~(\ref{Psieqn}) to spin-weighted spherical harmonics as follows:  Restoring 
the suppressed indices $s$, $\ell$, and $m$ to $S$ and $b_j$ in Eq.~(\ref{eq:SbY}), 
\be
   S = {}_sS_{\ell m}, \quad b_j = b_{\ell j m},
\ee
with $s=2$, we have 
\bea
\Psi &=& 8\sum_{\ell m}\frac{(-1)^mD\bar{R}_{\ell m}+12imM\Omega R_{\ell m}}{D^2+144M^2m^2\Omega^2}e^{im(\phi-\Omega t)}{\,}_2S_{\ell m}(\theta) \nonumber \\
&=& 8\sum_{\ell m}\frac{(-1)^mD\bar{R}_{\ell m}+12imM\Omega R_{\ell m}}{D^2+144M^2m^2\Omega^2}e^{-im\Omega t}\sum_{\ell^\prime}b_{\ell \ell' m}{\,}_2Y_{\ell^\prime,m}(\theta,\phi) \nonumber \\
&=& \sum_{\ell^\prime,m}\left(\sum_{\ell}8b_{\ell \ell^\prime m}\frac{(-1)^mD\bar{R}_{\ell m}+12imM\Omega R_{\ell m}}{D^2+144M^2m^2\Omega^2}e^{-im\Omega t}\right) {\,}_2Y_{\ell^\prime,m}(\theta,\phi) \nonumber \\
&=& \sum_{\ell m}\Psi_{\ell m}(t,r){\,}_2Y_{\ell m}(\theta,\phi).
\eea
%CHANGE OVERALL SIGN
To write the nonzero tetrad components of the metric perturbation of Eq.~(\ref{eq:hab}), we replace 
spin-coefficients by their values for Kerr given in Eq.~(\ref{eq:spincoeff}), we write the 
derivatives operators $\bm\Delta$ and $\bm D$ in the explicit forms implied by (\ref{eq:tetrad}), 
and we use Eq.~(\ref{green_eth_bar_b}) to replace 
the angular derivative operator $\bm\delta$ by its form in terms of $\eth$. We thereby obtain 
\begin{align}
h_{\bm{11}} &= \sum_{\ell m} \frac{1}{2\varrho^2} \bigl[ \bar{\eth}^2  %changed
        - 2ia(\varrho+im\Omega)\sin\theta\ \bar{\eth} - 3a^2\varrho^2\sin^2\theta
        + a^2(\varrho-im\Omega)(3\varrho+im\Omega)\sin^2\theta \bigr] \Psi_{\ell m}{\,}_2Y_{\ell m} + \rm{c.c.},\quad\\
%\end{align}
%\begin{align}
h_{\bm{13}} &= \sum_{\ell m} \frac{-1}{4\sqrt2}\frac{\Delta}{\Sigma\varrho^3} \Biggl[ 2\partial_r \bar{\eth} %changed
                -i a\left(\varrho-\bar\varrho+2im\Omega -2i \frac{a}\Sigma\cos\theta\right)\sin\theta \partial_r 
        - 2\left(\varrho+\bar\varrho + 4 \frac{r-M}\Delta +iq \right)\bar{\eth}
\nonumber \\
&\hskip3cm +6ia\varrho^2\sin\theta - ia\left(2\varrho+\bar{\varrho}-im\Omega+2i \frac{a}{\Sigma}\cos\theta\right) 
        \left(3\varrho + 4\frac{r-M}{\Delta}+iq\right)\sin\theta \nonumber \\
&\hskip7
cm -ia\left(3\varrho+im\Omega \right)\left(\varrho-2\bar\varrho-4\frac{r-M}\Delta -iq \right)\sin\theta \Biggr]\Psi_{\ell m}{\,}_2Y_{\ell m},\\
%\end{align}
%\bea
h_{\bm{33}} &= \sum_{\ell m} \frac14\frac{\Delta^2}{\Sigma^2\varrho^4}\Biggl[ \partial_r^2 + \left(\varrho-\bar{\varrho}+8\frac{r-M}{\Delta}-\frac{2r}{\Sigma}+2iq\right)\partial_r + 3\varrho^2 + \frac{4}{\Delta} - \frac{8(r-M)^2}{\Delta^2} +2i\frac{q(r-M)+m\Omega r}\Delta\nonumber \\ %changed
& \hskip5.5cm+ \left(2\varrho + \bar{\varrho}-4\frac{r-M}{\Delta} +2\frac{r}{\Sigma}+2iq\right) 
          \left(3\varrho + 4\frac{r-M}{\Delta} +iq\right) \Biggr]\Psi_{\ell m}{\,}_2Y_{\ell m}.
%\eea
\end{align}
Because $\Psi$ has spin-weight 2 and $\bar\eth\Psi$ has spin-weight 1, 
$\bar\eth$ acting on $\Psi$ and on $\bar\eth\Psi$ has the form  
(\ref{green_eth_bar_b}) with $s=2$ and $s=1$, respectively.

 We next expand the operators acting on $\Psi$ in powers of the small parameter $\cos\theta$  
and show below that terms involving $\cos\theta$ and $\cos^2\theta$ do not contribute to  
$h^{\rm ren}_{\bm\mu\bm\nu}$ (or to the nonzero part of the singular field).
Using Eq.~(\ref{green_sylm_a}) for the action of $\bar\eth$ on ${\,}_2Y_{\ell m}$, we obtain

\bea
h_{\bm{11}} &=& \sum_{\ell m}\Psi_{\ell m} \left( \sum_{n,s=0}^2{\,}_s{\cal A} ^n_{\ell m}\cos^n\theta{\,}_sY_{\ell m}\right) + \rm{c.c.},
\label{eq:h11}\eea
where the radial functions ${\,}_s{\cal A} ^n_{\ell m}$ are given by
\bsube\begin{align} %changed 
{\,}_0{\cal A} ^0_{\ell m} &= \sqrt{\frac{(\ell+2)!}{(\ell-2)!}}\,\frac{r^2}{2}, \\
{\,}_1{\cal A} ^0_{\ell m} &= -\sqrt{(\ell-1)(\ell+2)}\, ar(i+m\Omega r),  \\
{\,}_2{\cal A} ^0_{\ell m} &=  \frac12 a^2m\Omega r(2i+m\Omega r),  \\
{\,}_0{\cal A} ^1_{\ell m} &= - i\sqrt{\frac{(\ell+2)!}{(\ell-2)!}}\, ar ,  \\ 
{\,}_1{\cal A} ^1_{\ell m} &= -\sqrt{(\ell-1)(\ell+2)}\, a^2(1-2im\Omega r),  \\
{\,}_2{\cal A} ^1_{\ell m} &=  m\Omega a^3 (1-imr\Omega),  \\
{\,}_0{\cal A} ^2_{\ell m} &= -\frac1{2}\sqrt{\frac{(\ell+2)!}{(\ell-2)!}}\,a^2,  \\
{\,}_1{\cal A} ^2_{\ell m} &= \frac12 \sqrt{(\ell-1)(\ell+2)}\, a[ir+m\Omega(r^2+2a^2)],  \\
{\,}_2{\cal A} ^2_{\ell m} &= -\frac12 m\Omega a^2[2ir+m\Omega(r^2+a^2)];
\end{align}
\esube
\begin{align}
h_{\bm{13}} = \sum_{\ell m} \sum_{s=1}^2\sum_{n=0}^2 {\,}_s{\cal B} ^n_{\ell m}\cos^n\theta{\,}_sY_{\ell m},
\label{eq:h13}\end{align}
where, with $q$ defined by Eq.~(\ref{eq:q}), the functions ${\,}_s{\cal B} ^n_{\ell m}$ are given by
\bsube\begin{align} %changed
{\,}_1{\cal B} _{\ell m}^0 &= -\sqrt{\frac{(\ell-1)(\ell+2)}8}
        \left[ r\Delta\Psi_{\ell m}^\prime + (2r^2-2a^2-iqr\Delta)\Psi_{\ell m}\right] ,\\
%\end{align}
%%%%%%%
%\begin{align} %changed
{\,}_2{\cal B} _{\ell m}^0 &= \frac1{\sqrt{8}}am\Omega
        [ r\Delta\Psi_{\ell m}^\prime + (2r^2-2a^2-iqr\Delta)\Psi_{\ell m}],\\
%\end{align}
%%%%%%%
%\begin{align} %changed
{\,}_1{\cal B} _{\ell m}^1 &= 
     3i\sqrt{\frac{(\ell-1)(\ell+2)}8}\,\frac ar 
        [ r\Delta\Psi_{\ell m}^\prime + (2r^2-2a^2-iqr\Delta)\Psi_{\ell m}], \\
%\end{align}
%%%%%%%f
%\begin{align} %changed
{\,}_2{\cal B} _{\ell m}^1 &= -\frac1{\sqrt8}\frac{a^2}{r} \Biggl\{ (2+3im\Omega r)\Delta\Psi'_{\ell m}
                +\left[ 8(r-M) +3m\Omega qr\Delta -2ima+4im\Omega(2r^2-a^2)\right]\Psi_{\ell m}\Biggr\},\\ 
%\end{align}
%%%%%%%
%\begin{align} %changed
{\,}_1{\cal B} _{\ell m}^2 &= \sqrt{\frac{(\ell-1)(\ell+2)}2}\frac{a^2}{r^2} 
        \left[2r\Delta\Psi_{\ell m}^\prime + ( 3r^2+2Mr-5a^2-2iqr\Delta)\Psi_{\ell m}\right] ,\\
%\end{align}
%%%%%%%
%\begin{align} %changed
{\,}_2{\cal B} _{\ell m}^2 &=\frac1{32}\frac a{r^2}\{ [m\Omega r(r^2+8a^2)+12ia^2]\Delta\Psi'_{\ell m} 
        +[12ma^3+2m\Omega(r^4-a^2r^2+4Ma^2r-16a^4)]\Psi_{\ell m}\nonumber \\ 
        &\hskip6.5cm +i[48a^2(r-M)-m\Omega qr(r^2+8a^2)\Delta]\Psi_{\ell m}\};
\end{align} \esube
and
\bea
h_{\bm{33}} &=& \sum_{\ell m}\sum_{n=0}^2 {\cal C} _{\ell m}^n \cos^n\theta{\,}_2Y_{\ell m},
\label{eq:h33}\eea
with ${\cal C} _{\ell m}^n$ given by
\bea %changed
{\cal C} _{\ell m}^0 &=& \frac14\Delta^2\Psi_{\ell m}^{\prime\prime} 
        + \frac12\frac\Delta r \left(3r^2-2Mr-a^2 - iqr\Delta\right)\Psi_{\ell m}^\prime \nonumber \\
&& + \Biggl[ r^2-2M^2-a^2+2\frac{Ma^2}{r}-\frac14 q^2\Delta^2 
        -\frac i2\frac{q\Delta}r(2r^2-Mr-a^2)\Biggr]\Psi_{\ell m},
\eea
%%%%%%
\bea %changed
{\cal C} _{\ell m}^1 &=& \frac{-ia\Delta^2}{r}\Psi_{\ell m}^{\prime\prime} - \frac{ia\Delta}{r}\Psi_{\ell m}^\prime\left( 8(r-M)-\frac{3\Delta}{2r}-2im[a-\Omega(r^2+a^2)]\right) \nonumber \\
&& - \frac{ia}{r}\Psi_{\ell m}\Biggl[ 2im\Omega\Delta r + 8(M-2)^2 -2\Delta + \frac{6M\Delta}{r} + 6im[a-\Omega(r^2+a^2)](M-r) \nonumber \\
&& + \frac{3\Delta im[a-\Omega(r^2+a^2)]}{2r} -m^2\left[a-\Omega(r^2+a^2)\right]^2\Biggr],
\eea
%%%%%%
\bea %changed
{\cal C} _{\ell m}^2 &=& \frac{-2a^2\Delta^2}{r^2}\Psi_{\ell m}^{\prime\prime}  - \frac{\Delta a^2}{r^2}\Psi_{\ell m}^\prime\Biggl[16(r-M) - \frac{5\Delta}{2r} - 4im[a-\Omega(r^2+a^2)] \Biggr] \nonumber \\
&& - \frac{a^2}{r^2}\Psi_{\ell m}\Biggl[ 4im\Omega\Delta r + 16(M-r)^2 - 2\Delta + \frac{10M\Delta}{r} + 12im(M-r)[a-\Omega(r^2+a^2)] \nonumber \\
&& + \frac{5\Delta[a-\Omega(r^2+a^2)]}{2r} - 2m^2[a-\Omega(r^2+a^2)]^2\Biggr].
\eea

We now argue that the terms 
involving $\cos^n\theta$ with $n\neq 0$ in Eqs.~(\ref{eq:h11}), (\ref{eq:h13}) 
and (\ref{eq:h33}) can be ignored.  We refer here to our description in  
Sec. \ref{sec:hrad}  of the the renormalization procedure in terms of 
$\psi_0$ and $\Psi$.  The components $h^{\rm ren}_{\bm{\mu\nu}}$ of the 
renormalized radiation-gauge metric are given by Eqs.~(\ref{eq:h11}), 
(\ref{eq:h13}) and (\ref{eq:h33}), with $\Psi$ replaced by $\Psi^{\rm ren}$.  
Because  $\Psi^{\rm ren}$ is smooth and hence finite at the particle, 
no term proportional to $\cos\theta$ contributes to $h^{\rm ren}_{\bm{\mu\nu}}$. \\ 

It can happen, however, that the $\cos\theta$ terms contribute 
to $h^{\rm S}_{\bm{\mu\nu}}$, written in terms of $\Psi^{\rm S}$.  
The $\cos^2\theta$ terms cannot contribute, because they are 
$O(\rho^2)$ and multiply terms whose sum is at most $O(\rho^{-1})$ (terms 
involving two derivatives of $\Psi$).    
Similarly, the $\cos\theta$ terms cannot contribute to the leading 
term in $h^S$, because they are one order in $\rho$ smaller than the 
leading term in $H^s$.  At subleading order, however, they give an 
$O(\rho^0)$ contribution that has odd parity.  The parity of this 
contribution to the singular field follows from that fact that, at leading order, 
$\Psi^{\rm ret}$ is even under parity about the position of the particle 
(shown in detail in Sec. IIID of \cite{sf2}), while $\cos\theta$ is odd.  
The leading-order contributions to the $\cos\theta$ terms come from 
terms involving two derivatives of $\Psi^{\rm ret}$, and these are again 
even under parity, implying that at leading order, the contribution 
from $r>r_0$ cancels the contribution with opposite sign from $r<r_0$.
Finally, lower-order contributions multiplying $\cos\theta$ are 
order $\rho$, vanishing at the particle.  

One can then compute $H^{\rm ren}$ by subtracting the leading part of the 
singular field (which coincides the Lorenz-gauge singular field) and
by omitting terms that involve $\cos\theta$. We have verified this agreement 
numerically (see Sec.~\ref{sec7}), finding  
that the order $L^0$ part 
of $H^{\rm S}_\ell$ agrees to one part in $10^{12}$ with its analytic form computed by 
Linz \cite{linz}  and that the $O(L^{-1})$ contribution 
vanishes to within the accuracy of the computation.

\section{Gauge-invariant quantities}
\label{sec5}

In this section, we obtain expressions for the related quantities $\Delta U$ and $\hat\Delta \Omega$ that give, respectively, the change in the redshift factor 
of a trajectory at fixed angular velocity and the change in the angular velocity of a trajectory at fixed redshift factor $U$.   Each of these quantities is invariant under gauge-transformations generated by helically symmetric gauge vectors $\xi^\alpha$ and each can be written in terms of the similarly gauge-invariant 
quantity 
\be 
\dis H^{\rm ren} :=\frac12 h^{\rm ren}_{\alpha\beta}u^\alpha u^\beta,
\label{H}\ee
where $h^{\rm ren}_{\alpha\beta}$ is the renormalized metric perturbation.   

As shown by Mino et al. \cite{MinoSasaki} (see also Quinn and Wald \cite{quinnwald} and Detweiler $\&$ Whiting \cite{Detweiler:2002mi}), 
at order ${\frak m}/M$
the particle moves along a geodesic of the metric $g_{\alpha\beta} + h^{\rm ren}_{\alpha\beta}$,
where $g_{\alpha\beta}$ is the background (Kerr) metric.  Denote by $\hat u^\alpha = U (t^\alpha+\Omega\phi^\alpha)$ 
the particle's 4-velocity, normalized with respect to $g_{\alpha\beta} + h^{\rm ren}_{\alpha\beta}$, 
\be
 (g_{\alpha\beta} + h^{\rm ren}_{\alpha\beta})\,\hat u^\alpha \hat u^\beta = 1.
\label{hatnorm}\ee  
We consider first the difference $\Delta U$ between the value of $U$ for a circular geodesic of 
the perturbed metric and its value at the circular geodesic of the unperturbed metric with the 
same value of angular velocity $\Omega$.   Formally, because perturbations to first-order in 
$\frak m/M$ are linear in $\frak m$, we can write 
\be 
   \Delta U= \frak m\frac\partial{\partial\frak m}U(\frak m,\Omega)|_{\frak m=0}.
\label{Delta}\ee  

We will denote by $\delta U$ the gauge-dependent change in $U$ at a fixed value of $r$. Denoting 
by $U(\frak m, r)$ value of $U$ for the circular geodesic at radial coordinate $r$ of the metric 
$g_{\alpha\beta} + h^{\rm ren}_{\alpha\beta}$, we have  
\be 
   \delta U :=  \frak m\frac\partial{\partial\frak m}U(\frak m, r)|_{\frak m=0}. 
\label{delta}\ee 
Let $\xi^\alpha$ be the radial vector joining the unperturbed circular geodesic with angular velocity 
$\Omega$ to the perturbed geodesic with the same angular velocity: Formally 
\be
   \Omega(\frak m=0,r) = \Omega(\frak m,r+\xi^r) + O(\frak m^2).
\ee 
Then 
\be 
   \Delta U = \delta U + \Lie_{\bm\xi} U,  \quad \Delta \Omega = \delta \Omega + \Lie_{\bm\xi} \Omega = 0,
\ee
where $\Delta \Omega$ and $\delta \Omega$ are defined as in Eqs.~(\ref{Delta}) and (\ref{delta}).

   We can now quickly compute $\Delta U$ from Eq.~(\ref{hatnorm}), showing as follows the relation 
\be 
   \Delta U = -u^t H^{\rm ren}.
\label{DeltaU}\ee 
Define $\tilde k^\alpha$ in the equatorial plane by  
$\tilde k^\alpha = t^\alpha + \Omega\phi^\alpha$, with $\Omega = \Omega(\frak m,r)$, and let $k^\alpha$ 
be the Killing vector $k^\alpha = t^\alpha + \Omega_0 \phi^\alpha$, where $\Omega_0$ is the 
angular velocity of the unperturbed orbit through $r=r_0$.  Then, applying $\Delta$ to the normalization 
equation (\ref{hatnorm}) and evaluating the expression at $r=r_0$, we have  
\be
   0 = \Delta(g_{\alpha\beta}U^2 \tilde k^\alpha \tilde k^\beta) =  (\delta + \Lie_{\bm\xi})(g_{\alpha\beta}U^2 \tilde k^\alpha \tilde k^\beta) = (h^{\rm ren}_{\alpha\beta}+ \Lie_{\bm\xi}g_{\alpha\beta}) u^\alpha u^\beta
	+ \frac2U\Delta U + 2U u_\alpha (\delta + \Lie_{\bm\xi})\tilde k^\alpha,  
\label{Deltanorm}\ee
with $\delta$ again the perturbation at fixed radius $r$, as in Eq.~(\ref{delta}). Using the fact that
$\xi^\alpha$ is helically symmetric, we now see as follows that the terms $\Lie_\xi g_{\alpha\beta}u^\alpha u^\beta$ and $(\delta + \Lie_{\bm\xi}) \tilde k^\alpha$ 
vanish.  At $r=r_0$, we have $U(\frak m=0, r_0) = u^t$, whence 
$u^\alpha = u^t k^\alpha$, and Eq.~(\ref{gaugeinv}) then implies 
$\Lie_\xi g_{\alpha\beta}u^\alpha u^\beta=0$.  Finally, because the coordinate system 
is independent of $\frak m$, we have $\delta t^\alpha = 0 = \delta\phi^\alpha$ 
(that is, $\bm\partial_t$ and $\bm\partial_\phi$ do not change), and the last term vanishes:   
\be 
(\delta + \Lie_{\bm\xi})\tilde k^\alpha|_{r=r_0} = \Lie_{\bm\xi} t^\alpha + \Omega_0\Lie_{\bm\xi}\phi^\alpha 
	= \Lie_{\bm\xi} k^\alpha = -\Lie_{\bm k}\xi^\alpha = 0.
\ee
From the two surviving terms on the right of Eq.~(\ref{Deltanorm}), we obtain the claimed form 
$\dis\Delta U = -\frac12 u^t h^{\rm ren}_{\alpha\beta}u^\alpha u^\beta$.
  
  The change in the angular velocity at fixed $U$ is similarly gauge invariant and is easily obtained from 
$\Delta U$.  With $\Omega$ regarded as a function of $\frak m$ and $U$ we define its change at fixed $U$ by 
\be 
\hat\Delta\Omega := \frak m\frac\partial{\partial\frak m}\Omega(\frak m,U)|_{\frak m=0}.
\ee
From the fact that, at fixed $\frak m$, $\Omega(\frak m, U)$ is the inverse of $U(\frak m, \Omega)$, it 
follows that 
\be
 \hat\Delta\Omega = - \frac\pa{\pa U} \Omega(\frak m=0,U) \Delta U, 
\ee 
implying 
\begin{align}
\hat\Delta\Omega = -\frac1{u_{\phi}u^t}H^{\rm ren}.
\end{align}
The resulting values of $\Delta U$ and $\hat{\Delta}\Omega$ are presented in Tables \ref{deltaU} and \ref{deltaOm}.

For completeness, we give here explicit expressions for the quantities $\delta U$ and $\delta\Omega$. 
These depend on the gauge-dependent acceleration $a^\alpha$, the self-force per unit mass, which is 
ordinarily defined with the perturbed trajectory parametrized by proper time with respect to the {\em background metric}, 
implying for the 4-velocity $u^\alpha$ the normalization 
\be
 g_{\alpha\beta}\,u^\alpha u^\beta = 1.
\label{uwm}\ee
Denoting by $\tau$ and $\hat\tau$ proper time along a trajectory with respect to $g_{\alpha\beta}$ and 
$g_{\alpha\beta}+ h^{\rm ren}_{\alpha\beta}$, respectively, we have 
$\dis u^\alpha = \frac{d\hat\tau}{d\tau}\hat u^\alpha$, with 
\begin{align}
\frac{d\tau}{d\hat{\tau}} = 1 - \frac{1}{2} h_{\alpha\beta}^{\rm ren}u^\alpha u^\beta = 1 - H^{\rm ren}.
\end{align}
The acceleration of the perturbed trajectory with respect to the background metric $g_{\alpha\beta}$ is given by 
\bsube\bea
a^\alpha&:=& u^\beta\nabla\upp0\,_\beta u^\alpha 
\label{accel}\\
 &=& - (g^{\alpha\delta}+u^\alpha u^\delta)(\nabla\upp0\,_\gamma h_{\beta\delta}-\frac12\nabla\upp0\,_\delta h_{\beta\gamma})u^\beta u^\gamma,
\eea\esube
where $\nabla\upp0\,_\alpha$ is the covariant derivative operator of $g_{\alpha\beta}$.  

Using Eqs. (\ref{uwm}) and (\ref{accel}), we find for the changes in $\Omega$ and $u^t$ of a trajectory at fixed radius 
$r_0$ 
\begin{align} \label{Omega}
\delta\Omega &= \Omega \frac{2aM^{1/2}r_0^2+r_0^{5/2}(r_0-3M)}{2M(r_0^{3/2}+aM^{1/2})}a_r
\end{align}
and
\begin{align}
\delta u^t &= u^t \frac{r_0^{1/2}(r_0^2+a^2-2aM^{1/2}r_0^{1/2})}{2(r_0^{3/2}+aM^{1/2})}a_r , 
\end{align}
where $\Omega$ and $u^t$ are given by the relations in Eq (\ref{utOmega}), $\Omega$ is the frequency measured by the observer at infinity.  Here one is comparing the values of $\Omega$ and $u^t$ for circular geodesics of the 
perturbed metric $g_{\alpha\delta}+h^{\rm ren}_{\alpha\beta}$ to their values for a circular geodesic of 
the unperturbed metric $g_{\alpha\beta}$ at the same value $r=r_0$ of the radial coordinate.  The expressions 
are valid in any gauge, but the values of $a_r$, $\delta \Omega$ and $\delta U$ are gauge-dependent.

%%%%%%%%%%%%%%%%%%%%%%%%%%%%%%%%%%%%%%%%%%%%%%%%%%%%%%%%%%%%%%%%%%%%%%%%

\section{Lower multipoles}
\label{sec6}
The metric recovered from $\psi_0^{\rm ren}$ specifies the perturbation up to 
the contribution that comes from the change in mass and angular momentum of a 
Kerr metric and from a change in the center of mass that is pure gauge except at 
$r=r_0$ (and that does not contribute to $H^{\rm ren}$).%
\footnote{In the mode-sum renormalization, the individual 
modes of the metric are computed as the limits of their values 
as $r \rightarrow r_0$ from $r < r_0$ or $r > r_0$. Because the 
because the metric perturbation associated with a change in the 
center of mass is pure gauge for $r\neq r_0$, 
these limits vanish.}
In this section we calculate 
the contribution to the gauge-invariant $H^{\rm ret}$ from the 
change in mass and angular momentum due to the presence of the orbiting particle of mass 
$\mathfrak{m}$. 
We calculate them in the ``Kerr gauge"; that is, they are written as the 
first-order perturbations of the Kerr metric in Boyer-Lindquist 
coordinates associated with the changes $\delta M$ and $\delta J$ in its mass
and angular momentum.  These two parts of the metric perturbation are thus stationary 
and axisymmetric, and they are associated with a stationary, axisymmetric part of the 
stress-energy tensor.  For a particle in circular orbit, $\delta M$ and $\delta J$ have the simple forms  
\be
 \delta M = E = mu_\alpha t^\alpha, 
\qquad
 \delta J = L = -m u_\alpha\phi^\alpha 
\ee  
(for our $+---$ signature), as stated by L. Price \cite{lpthesis}.  
The expression for $\delta J$ follows, for example, from the Komar formula for 
angular momentum, valid for a stationary axisymmetric perturbation of a stationary 
axisymmetric spacetime;  it implies 
\be
  \delta J = -\int_V T^\alpha_\beta\phi^\beta dS_\alpha = -mu_\phi,  
\label{dJ}\ee
when there is no change in the angular momentum of the black hole.  
The change in the mass follows from the Bardeen-Carter-Hawking first law of 
thermodynamics for black holes and matter, which gives 
\be
   \delta M = \Omega \delta J + \frak m/u^t = \frak m u_t. 
\ee
(One can also use the generalization of the first law to helically symmetric 
binaries \cite{fus,fuse,ltbw}, but the stationarity and 
axisymmetry of the relevant perturbation again means that the original 
form of the first law is valid.)  

Although one can also find $\delta M$ directly from the Komar expressions 
in terms of the timelike Killing vector shared 
by the background and perturbed spacetimes, a warning is needed:  An unexpected 
subtlety arises in using the Komar expression for the mass.  The Komar mass on a sphere 
outside the particle orbit has the form 
\be
  \delta M = \delta \frac1{4\pi} \int \nabla^\alpha t^\beta dS_{\alpha\beta},
\label{mk}\ee
where $dS_{\alpha\beta} = \frac12 \epsilon_{\alpha\beta\gamma\delta}dS^{\alpha\beta}$  
(implying $dS_{tr} = \frac12 \sqrt{-g} d\theta d\phi$).   
If one requires that the change in the Komar mass of the black hole vanish and assumes 
that $\delta t^\alpha =0$, the result is an incorrect expression for $\delta M$. 
Gauss's theorem gives   
\bea
  \delta M &=& \delta \frac1{4\pi} \int_V \nabla_\beta \nabla^\alpha t^\beta dS_{\alpha} 
		+ \delta \frac1{4\pi} \int_{\rm horizon} \nabla^\alpha t^\beta dS_{\alpha\beta} \nonumber\\
	   &=& \int (2T^\alpha{}_\beta - \delta^\alpha_\beta T) t^\beta dS_\alpha 
		+ \delta \frac1{4\pi} \int_{\rm horizon} \nabla^\alpha t^\beta dS_{\alpha\beta},
\eea
and 
\be
  \int (2T^\alpha{}_\beta - \delta^\alpha_\beta T) t^\beta dS_\alpha = \frak m (2u_t - 1/u^t)\neq \frak m u_t.
\ee
The discrepancy arises from a rescaling of $h_{tt}= h_{\alpha\beta} t^\alpha t^\beta$ near the horizon for a time-independent perturbation.
This is easiest to see for a Schwarzschild background, where the perturbation in the mass arises from the 
spherically symmetric part of the perturbation -- from the perturbation due to a spherical shell of 
dust whose particles have trajectories isotropically distributed over all circular 
geodesics at $r=r_0$.  The Komar 
mass is gauge invariant,% 
\footnote{Gauge invariance of the integral (\ref{mk}), over a sphere $S$ where there is 
no matter, can be seen as follows.  Let $S'$ be another sphere homologous to $S$ with
no matter in the region between them.  Then, for any given choice of gauge, the 
value of the Komar integral is the same on the two spheres. Consider a gauge transformation
associated with an arbitrary gauge vector $\xi^\alpha$ defined in a neighborhood of 
$S$, and extend $\xi^\alpha$ smoothly so that it vanishes on $S'$. In the new gauge, 
the value of the Komar integral on $S'$ has not changed; and it must again have the 
same value on $S$ and $S'$.  Its value on $S$ is therefore unchanged.}
and we can can compute it in a Schwarzschild gauge.  The perturbed 
field equation then requires continuity of $h_{tt} =: e^{2\Phi}$ across $r=r_0$, and $\delta\Phi$ 
is constant inside $r=r_0$, implying a constant rescaling of time for $r<r_0$. The result is that  
the expression for the change in the Komar mass at the horizon is evaluated with a rescaled metric 
but without a rescaled $t^\alpha$, giving a nonzero result,    
\be
 \delta \frac1{4\pi} \int_{\rm horizon} \nabla^\alpha t^\beta dS_{\alpha\beta}
	= \frak m (1/u^t - u_t),
\ee
that yields the correct value $\delta M = \frak m u_t$ for the change in the spacetime mass.  
The change in $h_{\alpha\beta}$ inside $r=r_0$ has the form $\Lie_{\bm\xi}g_{\alpha\beta}$ 
for a vector $\xi^\alpha$ linear in $t$; because $t^\alpha$ remains fixed, however, this is  not a gauge transformation of the integrand $\nabla^\alpha t^\beta dS_{\alpha\beta}$ (the integrand does not change 
by $\Lie_{\bm\xi} (\nabla^\alpha t^\beta dS_{\alpha\beta}) $).     

Finally, Eq.~(\ref{dJ}) for the change in angular momentum is 
valid, because the rescaling does not alter the Komar expression for the 
angular momentum at the horizon.

To calculate the metric perturbation that comes from the change in mass and angular momentum, we find the 
first order perturbation of the (relevant components of the) Kerr metric in Boyer-Lindquist coordinates 
which are
\begin{align}
h_{tt} &= -\frac{2\delta M}{r} \nonumber \\
h_{t\phi} &= 0 \nonumber \\
h_{\phi\phi} &=  \frac{2(M+r)a^2\delta M}{M r}
\end{align}
for the change in mass, and
\begin{align}
h_{tt} &= 0 \nonumber \\
h_{t\phi} &=  \frac{2M\delta J}{r} \nonumber \\
h_{\phi\phi} &= -\frac{2 a (2M+r)\delta J}{Mr}
\end{align}
for the change in angular momentum. From these expressions and Eqs.~(\ref{utOmega}) for $u^t$ and $\Omega$ 
(and using $u^\phi=\Omega u^t$), we obtain
\begin{align}
%H_{\delta M} = \frac{\frak{m}(r^{3/2}-2Mr^{1/2}\pm a M^{1/2})}{r^{1/4}(r^{3/2}-3Mr^{1/2}\pm 2aM^{1/2})^{3/2}}
H_{\delta M} = \frac{\frak{m}(r^2 \pm 2a M^{1/2}r^{1/2}-a^2)(r^{3/2}-2M r^{1/2}\pm a M^{1/2})}{r^{9/4}(r^{3/2}-3M r^{1/2}\pm2aM^{1/2})^{3/2}}
\end{align}
and
\begin{align}
H_{\delta J} = \frac{ M^{1/2}\frak{m}(r^2\mp2aM^{1/2}r^{1/2}+a^2)(\pm a - 2M^{1/2}r^{1/2}) }{ r^{9/4}(r^{3/2}-3Mr^{1/2}\pm 2aM^{1/2})^{3/2} }
\end{align}
where the upper (lower) sign is used for direct (retrograde) orbits.
%%%%%%%%%%%%%%%%%%%%%%%%%%%%%%%%%%%%%%%%%%%%%%%%%%
%%%%%%%%%%%%%%%%%%%%%%%%%%%%%%%%%%%%%%%%%%%%%%%%%%

\section{Numerical results}
\label{sec7}

The renormalization of $H$ follows the mode-sum method described in Sec.~\ref{sec:hrad}.
After the odd-parity terms -- terms involving $\cos\theta$ in Eqs.~(\ref{eq:h11}), (\ref{eq:h13}) 
and (\ref{eq:h33}) -- are omitted, the method is identical to that 
used in \cite{sf3} for a particle in circular orbit in a Schwarzschild background.
 
In Eq.~(\ref{eq:hsr}), with the odd-parity part of $H^{\rm S}$ gone, the remaining 
$O(L^{-2})$ terms vanish at the particle, allowing us to write the remaining part $\widetilde H_\ell^{\rm S}$ of $H_\ell^{\rm S}$ in the form 
\begin{align}
\widetilde H_{_\ell}^{\rm S} = E_0 + \sum_{k=1}^{k_{\rm max}} \frac{E_{2k}}{P_{2k}(\ell)},
\end{align}
where $P_{2k}(\ell)$ is a polynomial in $\ell$ 
of order $2k$ for which
\begin{align}
\sum_{\ell=0}^{\infty} \frac{1}{P_{2k}(\ell)} = 0.
\end{align} 
We numerically match $\widetilde H_{\ell}^{\rm ret}$ (where the tilde again denotes 
a value computed with odd-parity terms omitted) to this expansion of 
$\widetilde H_{\ell}^{\rm S}$,
\begin{align}
\widetilde H_{_\ell}^{\rm S} = E_0 + \sum_{k=1}^{k_{\rm max}} \frac{E_{2k}}{P_{2k}(\ell)},
\end{align}
and extract the regularization coefficients $E_{2k}$ up to $k_{\rm max}$ between 8 and 10.    
The method used in the numerical matching and an error-minimization criterion for the 
choice of $k_{\rm max}$ are described in detail in \cite{sf3}.  
The resulting value of $H^{\rm ren}$ is given by
\begin{align}
H^{\rm ren} = \sum_{\ell=0}^{\ell_{\rm max}}\left[ H_{\ell}^{\rm ret} -  H_{\ell}^{\rm S} \right]
	= \sum_{\ell=0}^{\ell_{\rm max}}\left[ \widetilde H_{\ell}^{\rm ret} - \widetilde H_{\ell}^{\rm S} \right],
\end{align}
with $\ell_{\rm max} = 74$.
The analytical value of $E_0$ \cite{linz} is given by
\begin{align}
E_{0 {\rm analytical}} = \frac{2}{\pi\sqrt{(1+\beta )g_{\theta\theta}}}K \left( \frac{\beta}{1+\beta} \right),
\end{align}
where $\beta = \frac{ g_{\phi\phi} - g_{\theta\theta} + L^2 }{ g_{\theta\theta} }$ and $K$ is the complete elliptic integral of the first kind, $\dis K(m)=\int_0^{\pi/2}(1-m\sin^2\phi)^{-1/2} d\phi$. We compare the value of $E_0$ obtained by numerical matching to the above analytical result and observe that they agree to 12 significant figures.

\section{Discussion and future work}

The results here are based on the computation of the invariant $H^{\rm ren}$, and work now underway with 
A. Le Tiec shows that $\Delta U$ from the computations in our modified radiation gauge agrees with the post-Newtonian series for $\Delta U$ linear in the spin parameter $a/M$: A preliminary matching shows that the first coefficient in the pN series agrees to five significant digits.  The results here also agree with those of a separate EMRI computation by Dolan, who works in a Lorenz gauge using an effective source method (agreement is within their numerical error bars of order $10^{-2}$). Finally, we have also begun work to extend the computation reported here to find the self-force on a particle in circular orbit in a Kerr background.

\begin{center}
\begin{table}
  \begin{tabular}{|c|c|c|c|c|c|c|c|c|}
    \hline
                 \multicolumn{1}{|p{0.7cm}|}
                 {\centering $r_0/M$}
                & \multicolumn{1}{|p{2.0cm}|}
                 {\centering $a=-0.9M$}
                & \multicolumn{1}{|p{2.0cm}|}
                 {\centering $a=-0.7M$}
                 & \multicolumn{1}{|p{2.0cm}|}
                 {\centering $a=-0.5M$}
                 & \multicolumn{1}{|p{2.0cm}|}
                 {\centering $a=0.0M$}
                 & \multicolumn{1}{|p{2.0cm}|}
                 {\centering $a=0.5M$}
                 & \multicolumn{1}{|p{2.0cm}|}
                 {\centering $a=0.7M$} 
                 & \multicolumn{1}{|p{2.0cm}|}
                 {\centering $a=0.9M$} \\

    \hline
       4        &                -                        &                -                        &                -                        &                -                       &                -                        &        -0.39639405           &        -0.32811192                \\
       5        &                -                        &                -                        &                -                        &                -                       &        -0.31443977          &        -0.27861234           &        -0.25156061                \\
       6        &                -                        &                -                        &                -                        &        -0.29602751         &        -0.23463184          &        -0.21756347           &        -0.20361838                \\
       7        &                -                        &                -                        &                -                        &        -0.22084753         &       -0.18875155          &        -0.17902998           &        -0.17073001                \\
       8        &                -                        &                -                        &        -0.20415909          &        -0.17771974         &       -0.15838853          &       -0.15222199           &       -0.14680897                \\
       10        &       -0.15129436        &        -0.14557511          &        -0.14033900          &        -0.12912227         &        -0.12019572          &        -0.11717475           &        -0.11443451                \\
       15        &      -0.083291764     &       -0.081933637        &        -0.080646922       &        -0.077725319       &        -0.075195106        &        -0.074284771        &        -0.073429473       \\
       20        &        -0.058142984      &        -0.057590366        &       -0.057059948       &        -0.055827719       &        -0.054723506        &       -0.054316065        &        -0.053927537        \\
       30        &        -0.036504919      &        -0.036334869        &       -0.036169772       &        -0.035778314       &        -0.035416550        &        -0.035279964        &        -0.035147937        \\
       50        &       -0.021026283      &        -0.020984416        &       -0.020943414       &        -0.020844656       &        -0.020751199        &        -0.020715285        &        -0.02068020        \\
       70        &         -0.014784459      &       -0.014767331        &        -0.014750491      &        -0.014709646       &        -0.014670583        &       -0.014655454       &        -0.014640606        \\
       100      &        -0.010234918      &        -0.010228170        &        -0.010221515       &        -0.010205282       &        -0.010189625        &       -0.010183523        &      -0.010177512        \\
    \hline
  \end{tabular}
\caption{This table presents the numerical values of $\Delta U$ for different values of $r_0/M$ and $a$. They are accurate to a fractional difference of order $10^{-8}$.}
 \label{deltaU}
  \end{table}
\end{center}

\begin{center}
\begin{table}
  \begin{tabular}{|c|c|c|c|c|c|c|c|c|}
    \hline
                 \multicolumn{1}{|p{0.7cm}|}
                 {\centering $r_0/M$}
                & \multicolumn{1}{|p{2.0cm}|}
                 {\centering $a=-0.9M$}
                & \multicolumn{1}{|p{2.0cm}|}
                 {\centering $a=-0.7M$}
                 & \multicolumn{1}{|p{2.0cm}|}
                 {\centering $a=-0.5M$}
                 & \multicolumn{1}{|p{2.0cm}|}
                 {\centering $a=0.0M$}
                 & \multicolumn{1}{|p{2.0cm}|}
                 {\centering $a=0.5M$}
                 & \multicolumn{1}{|p{2.0cm}|}
                 {\centering $a=0.7M$} 
                 & \multicolumn{1}{|p{2.0cm}|}
                 {\centering $a=0.9M$} \\

    \hline
       4        &                -                        &                -                         &                -                         &                -                        &                -                        &        0.054267340            &        0.052559297                \\
       5        &                -                        &                -                         &                -                         &                -                        &       0.047924050         &        0.046963137            &        0.046465534                \\
       6        &                -                        &                -                         &                -                         &        0.042727891          &        0.040850942        &        0.040470951            &        0.040275775               \\
       7        &                -                        &                -                         &                -                         &        0.036056740          &        0.035175043        &        0.034994056            &        0.034900960                \\
       8        &                -                        &                -                         &        -0.031876878         &        0.031046361          &        0.030576073        &        0.030478282            &       0.030427413                \\
       10      &       -0.024543706        &        -0.024365158        &        -0.024209291         &        0.023913779          &        0.023742658        &        0.023706031            &        0.023686199                \\
       15      &        -0.014462899        &        -0.014434162        &       -0.014408846         &        0.014359915          &       0.014330238        &        0.014323381            &        0.014319216                \\
       20      &       -0.0098123143      &       -0.0098041785     &        -0.0097969635        &        0.0097828022        &        0.0097738694     &        0.0097716692        &       0.0097702125       \\
       30      &        -0.0055825069      &        -0.0055810957     &       -0.0055798309        &        0.0055772872        &        0.0055755838     &        0.0055751254        &        0.0055747889        \\
       50      &        -0.0026871991      &       -0.0026870390     &        -0.0026868933        &        0.0026865907        &        0.0026863727     &        0.0026863081        &        0.0026862560        \\
       70      &        -0.0016464854      &        -0.0016464466     &        -0.0016464110        &        0.0016463355        &       0.0016462787     &        0.0016462609        &        0.0016462460        \\
       100    &        -0.00097498493   &       -0.00097497623   &        -0.00097496816      &        0.00097495060      &       0.00097493692  &        0.00097493244        &       0.00097492852        \\
    \hline
  \end{tabular}
\caption{Numerical values of $M\hat\Delta \Omega$ for different values of $r_0/M$ and $a$. 
The values are accurate to a fractional difference of order $10^{-8}$.}
 \label{deltaOm}
  \end{table}
\end{center}

\begin{acknowledgments}
We thank Sam Dolan and Alexandre Le Tiec for providing results of their work prior to 
publication, leading to corrections in our computations.  We also thank Leor Barack 
for helpful conversations.  This work was supported in part by NSF Grants PHY 0503366 
and PHY 1001515 and by European Research Council Starting Grant No. 202996.  

\end{acknowledgments}

\bibliography{sf4}
\end{document}